\documentclass[pdflatex, sn-standardnature]{sn-jnl}

\jyear{2022}%

\usepackage{hyperref}
\usepackage{subcaption}
\usepackage{mathtools}
\usepackage{amsmath}
\usepackage{nameref}
\usepackage[final]{pdfpages}

\newcommand{\stellarPAnum}{2039} 
\newcommand{\gasPAnum}{1503} 
\newcommand{\stellargasPAnum}{1310} 
\newcommand{\stellargasExcNum}{1091} 
\newcommand{\earlyPA}{423} 
\newcommand{\latePA}{864} 
\newcommand{\noclassPA}{23} 
\newcommand{\earlyPAfrac}{32\%} 
\newcommand{\latePAfrac}{66\%} 
\newcommand{\earlyPAExcfrac}{25\%} 
\newcommand{\latePAExcfrac}{73\%} 
\newcommand{\BPTnum}{1819} 
\newcommand{\misalignedFracThirty}{13$\pm1\%$} 
\newcommand{\misalignedFracForty}{10$\pm1\%$} 
\newcommand{\misalignedFracEarlyFortyMyCut}{25$\pm2\%$} 
\newcommand{\misalignedFracLateFortyMyCut}{3$\pm1\%$} 
\newcommand{\misalignedFracEarlyThirtyMyCut}{28$\pm2\%$} 
\newcommand{\misalignedFracLateThirtyMyCut}{6$\pm1\%$} 
\newcommand{\misalignedFracEarlyThirtyBryant}{39$\pm3\%$} 
\newcommand{\misalignedFracLateThirtyBryant}{6$\pm1\%$} 
\newcommand{\BPTAGN}{89}
\newcommand{\BPTLINER}{231}
\newcommand{\BPTSTAR}{1499}
\newcommand{\BPTAGNFrac}{5}
\newcommand{\BPTLINERFrac}{13}
\newcommand{\BPTSTARFrac}{82}

\newcommand{\misalignedBPTForty}{9$\pm$1\%}

\newcommand{\AGNalignedForty}{7$\pm$1\%} 
\newcommand{\AGNmisalignedForty}{20$_{-3}^{+5}$\%} 
\newcommand{\LINERalignedForty}{15$\pm$1\%} 
\newcommand{\LINERmisalignedForty}{43$\pm$5\%} 
\newcommand{\STELLARalignedForty}{78$\pm$1\%} 
\newcommand{\STELLARmisalignedForty}{37$\pm$5\%} 
\newcommand{\totalAGNalignedfrac}{77}

\newcommand{\blAGN}{17}
\newcommand{\blAGNBPTAGN}{12}
\newcommand{\blAGNBPTSF}{2}
\newcommand{\blAGNBPTLINER}{3}

\newcommand{\AGNalignedThirty}{7$\pm$1\%} 
\newcommand{\AGNmisalignedThirty}{17$_{-3}^{+3}$\%} 
\newcommand{\LINERalignedThirty}{16$\pm$1\%} 
\newcommand{\LINERmisalignedThirty}{35$\pm$4\%} 
\newcommand{\STELLARalignedThirty}{78$\pm$1\%} 
\newcommand{\STELLARmisalignedThirty}{49$\pm$4\%} 

\newcommand\aj{Astronomical Journal}

\newcommand\araa{Annual Review of Astron and Astrophys}
\newcommand\apj {Astrophysical Journal}
\newcommand\apjl{Astrophysical Journal, Letters}
\newcommand\apjs{Astrophysical Journal, Supplement}
\newcommand\apss{Astrophysics and Space Science}
\newcommand\aap{Astronomy and Astrophysics}
\newcommand\aapr{Astronomy and Astrophysics Reviews}
\newcommand\aaps{Astronomy and Astrophysics, Supplement}

\newcommand\mnras{Monthly Notices of the RAS}

\newcommand\pasa{Publications of the Astron. Soc. of Australia}
\newcommand\pasp{Publications of the ASP}

\newcommand\nat{Nature}

\usepackage{verbatim}

\begin{document}

\title[An increase in black hole activity in galaxies with misaligned gas]{An increase in black hole activity in galaxies with kinematically misaligned gas}

\author*[1,2,3]{\fnm{Sandra I.} \sur{Raimundo}}\email{s.raimundo@soton.ac.uk}

\author[1]{\fnm{Matthew} \sur{Malkan}}
\equalcont{These authors contributed equally to this work.}

\author[2,4]{\fnm{Marianne} \sur{Vestergaard}}
\equalcont{These authors contributed equally to this work.}

\affil*[1]{\orgdiv{Department of Physics and Astronomy}, \orgname{University of California}, \orgaddress{\city{Los Angeles}, \postcode{90095}, \state{California}, \country{USA}}}

\affil[2]{\orgdiv{DARK, Niels Bohr Institute}, \orgname{University of Copenhagen}, \orgaddress{\street{Jagtvej 155}, \city{Copenhagen N}, \postcode{2200}, \country{Denmark}}}

\affil[3]{\orgdiv{Department of Physics \& Astronomy}, \orgname{University of Southampton}, \orgaddress{\street{Highfield}, \city{Southampton}, \postcode{SO17 1BJ}, \country{UK}}}

\affil[4]{\orgdiv{Steward Observatory}, \orgname{University of Arizona}, \orgaddress{\street{933 N. Cherry Avenue}, \city{Tucson}, \postcode{AZ 85721}, \country{USA}}}

\abstract{External accretion events such as a galaxy merger or the accretion of gas from the immediate environment of a galaxy, can create a large misalignment between the gas and the stellar kinematics. Numerical simulations have suggested that misaligned structures may promote the inflow of gas to the nucleus of the galaxy and the accretion of gas by the central supermassive black hole. 
We show for the first time that galaxies with a strong misalignment between the ionised gas and stellar kinematic angles have a higher observed fraction of active black holes than galaxies with aligned rotation of gas and stars. 
The increase in black hole activity suggests that the process of formation and/or presence of misaligned structures is connected with the fuelling of active supermassive black holes.}

\keywords{Active Galactic Nuclei, supermassive black holes, galaxy}

\maketitle

\section{Introduction}
\label{sec:introduction}

Galaxy interactions are a predicted and observed feature of galaxy evolution (\cite{zwicky56}, \cite{haynes84}, \cite{binney&tremainebook}). This process often leaves an imprint that is observable in the aftermath of the interaction, such as tidal features or perturbations in the stellar and gas kinematics of the galaxy (e.g. \cite{haynes84},\cite{bertola92},\cite{sancisi08}). One of these perturbations is a strong misalignment between the kinematic position angle of the stars (PA$_{\rm stellar}$) and the gas (PA$_{\rm gas}$), defined as the orientation of the mean stellar and gas motions on a map of velocities (see \nameref{sec:methods} for more details). In some cases the misalignment can be $\Delta$PA $=$  $\vert$PA$_{\rm stellar}$ - PA$_{\rm gas}\vert$ = 180$^{\circ}$ which causes the visually striking feature of stars and gas rotating in opposite directions with respect to each other i.e. stellar-gas counter-rotation. This feature reflects the opposite angular momentum that the gas has with respect to the main stellar body of the host galaxy (e.g. \cite{kannapan&fabricant01}). It is inherently difficult to produce a large kinematic misalignment using internal processes in the galaxy (e.g. \cite{davies14}, \cite{raimundo17} and references therein), which indicates that a large misalignment ($\Delta$PA $\gtrsim$ 30$^{\circ}$) is a clear signature of a past external interaction, such as a major galaxy merger, minor merger or late-stage gas accretion, such as the infall of gas from a neighbour galaxy or  gas accretion triggered by a flyby \cite{bertola92}\cite{davis&bureau16}.

The identification of galaxies with misalignments has accelerated with the advent of galaxy surveys using integral field spectroscopy (IFS) (e.g. \cite{sarzi06}, \cite{garcia-lorenzo15}, \cite{jin16}, \cite{bryant19}). The two-dimensional maps of gas and stellar velocity produced by integral field spectrographs allows for a more unbiased identification of kinematic misalignments than the use of long-slit spectroscopy, especially for complex gas distributions. 
Several galaxy surveys using IFS have identified a substantial fraction of misalignment. For example, out of the slow- and fast-rotating early-type galaxies with ionised gas detections in the ATLAS 3D sample, 41$\%$ (55/133) show gas that is misaligned by more than 30 degrees with respect to the stars \cite{davis11}. In late-type galaxies large misalignments are less common, with $<$ 12\% of spirals showing counter-rotating ($\Delta$PA $\sim$ 180$^{\circ}$) gas discs \cite{pizzella04}. Large scale simulations show that counter-rotating discs can be long-lived, displaying stellar-gas counter-rotation for more than 2 Gyr after their formation \cite{starkenburg19}.

There is growing evidence from observational studies of single galaxies with misalignment (e.g. \cite{raimundo13}, \cite{raimundo17}, \cite{gnilka20}, \cite{raimundo21}) that the process of external gas accretion may drive gas to the nuclei of galaxies. A supply of nuclear gas that can be accreted by a central supermassive black hole is believed to be essential for the fuelling and powering of Active Galactic Nuclei (AGN) \cite{shlosman90}. \cite{davies14} suggested that black holes in S0 early-type galaxies are fuelled by externally accreted gas, a hypothesis that is supported by the (20 - 40\%) of S0 galaxies with gas that show counter-rotation (e.g. \cite{bertola92}, \cite{pizzella04}, \cite{katkov14}), pointing towards an external origin for most, if not all, of the gas in S0s. Considering that external gas accretion can also result in aligned stellar-gas kinematics (depending on the geometry of the interaction and galaxy morphology), the large fraction of early-type galaxies with misalignments points towards a major contribution of external accretion to the total gas content in those galaxies \cite{davis11}, and therefore to the gas available for black hole fuelling.

Theoretical studies have suggested that the presence of counter-rotating or significantly misaligned structures promote gas inflow (e.g. \cite{thakar&ryden96}, \cite{vandevoort15}) and potentially the fuelling of supermassive black holes in galaxies of all types (e.g. \cite{negri14}, \cite{capelo&dotti17}, \cite{taylor18}). However, this hypothesis has not been tested observationally. Considering the large fraction of galaxies that undergo external interactions at high and low redshift (e.g. \cite{conselice22}) and their potential consequences for black hole activity and star formation (e.g. \cite{dimatteo07}, \cite{silverman11}, \cite{kaviraj14}) understanding the impact of external accretion and misalignment to the fuelling of the black hole may shed light on the process of black hole accretion and activation. In this work we investigate for the first time the incidence of active supermassive black holes (AGN) in galaxies with and without an observed misalignment between gas and stars.

\section{Results}
\label{sec:results}

\begin{figure*}[h]
	\centering
	\includegraphics[width=12cm]{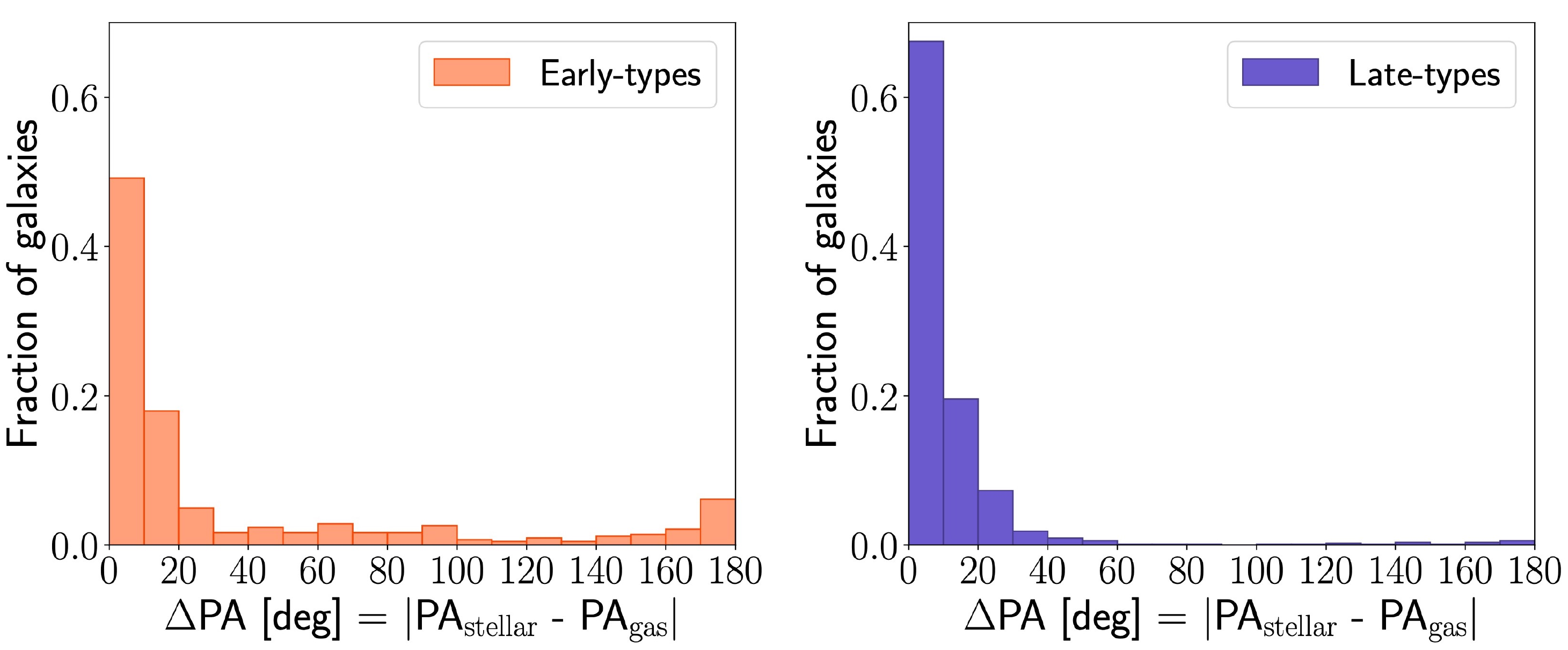}
	\caption{Distribution of the stellar to gas kinematic misalignment ($\Delta$PA = $\vert$PA$_{\rm stellar}$ - PA$_{\rm gas}\vert$) for the entire sample with measured PAs and morphology classification (\stellargasPAnum\ galaxies), divided into \earlyPA\ early-type galaxies (left panel) and \latePA\ late-type galaxies (right panel). In this sample, \noclassPA\ galaxies did not have a morphology classification. Late type galaxies mostly have aligned PAs ($\Delta$PA $\sim$ 0), while early-type galaxies have a higher percentage of their population spread out in $\Delta$PA.}
	\label{Fig1_main}
\end{figure*}

To investigate the number of AGN as a function of stellar to gas misalignment, we used a large sample of 3068 galaxies with redshifts 0.004 $<$ z $<$ 0.095 from the SAMI IFS survey \cite{croom21}, with available two-dimensional maps of stellar and ionised gas velocity. We determine the stellar and ionised gas kinematic position angles (PA$_{\rm stellar}$ and PA$_{\rm gas}$ respectively) for each galaxy, and the difference between these two angles which we refer to as misalignment: ($\Delta$PA = $\vert$PA$_{\rm stellar}$ - PA$_{\rm gas}\vert$) (see \nameref{sec:methods}). An example of this analysis can be found in Fig.~\ref{FigA1} of \nameref{sec:A1}. In Fig.~\ref{Fig1_main} we show the distribution of $\Delta$PA angles for the sample, divided according to galaxy morphology. Both early- and late-type galaxies have $\gtrsim 50\%$ of their population with well aligned stellar and gas kinematic angles (0$^{\circ} \leq \Delta$PA $< 10^{\circ}$). The remaining fraction of galaxies has $\Delta$PA spread out in the entire $\Delta$PA range, with a small but noticeable increase for counter-rotation ($\Delta$PA $\sim 180^{\circ}$). 

To separate the sample into `aligned' and `misaligned' galaxies we adopt the following classification: aligned galaxies (0$^{\circ} \leq \Delta$PA $<$ 45$^{\circ}$), misaligned galaxies (45$^{\circ} \leq \Delta$PA $\leq$ 180$^{\circ}$).
Since our main goal is to study the properties of misaligned galaxies, we take a more conservative approach to separate the two populations and adopt a threshold angle of 45$^{\circ}$ which is higher than the $\Delta$PA = 30$^{\circ}$ commonly used in the literature (e.g. \cite{davis11}). We use this higher value to minimise the contamination from galaxies that are aligned but may have small scale turbulence or kinematic deviations, and to still ensure a significant number of galaxies in each group. To ease comparison with other studies, we also quote the results found by using $\Delta$PA = 30$^{\circ}$ as a threshold angle in Fig.~\ref{FigA2} of \nameref{sec:A1}. 

Out of the \stellargasPAnum\ galaxies for which $\Delta$PA can be accurately determined, we find that \misalignedFracForty\ have a large misalignment ($\Delta$PA $\geq 45^{\circ}$). When dividing the sample according to morphology, we find that \misalignedFracEarlyFortyMyCut\ of the early-type galaxies (elliptical and S0 galaxies) have a misalignment of $\Delta$PA $\geq 45^{\circ}$ compared with only \misalignedFracLateFortyMyCut\ of the late-type galaxies (spirals), consistent with previous work \cite{bryant19}.  Misaligned galaxies tend to be relatively more common in the early-type galaxy population. 

\subsection{AGN identification}
To identify AGN in the sample we created spatially resolved Baldwin, Phillips \& Terlevich (BPT) diagrams (\cite{baldwin81}, \cite{veilleux&osterbrock87}) from the two-dimensional flux maps of several optical emission lines (\cite{green18}, \cite{medling18}), which allow us to do a spaxel by spaxel analysis of the full cube and separate spatial regions in our galaxies with gas excitation by young stars (HII regions with star formation), AGN or LINERs (Low Ionization Nuclear Emission Line Regions). We then classify our galaxies based on the fraction of spaxels in each region of the BPT diagrams - see \nameref{sec:methods}. It is known that optical AGN classification based on narrow emission lines may miss AGN populations (e.g. \cite{elvis81}). To identify additional AGN we searched for the presence of broad emission lines in the nuclear spectrum. We also cross-matched our galaxy catalogue with known multi-wavelength AGN catalogues in other wavebands (infrared, X-rays and radio). Table~\ref{Table1} in the \nameref{sec:A1} shows a summary of the number of galaxies and AGN in our sample.

\begin{figure}[h]
	\centering
	\includegraphics[width=0.8\linewidth]{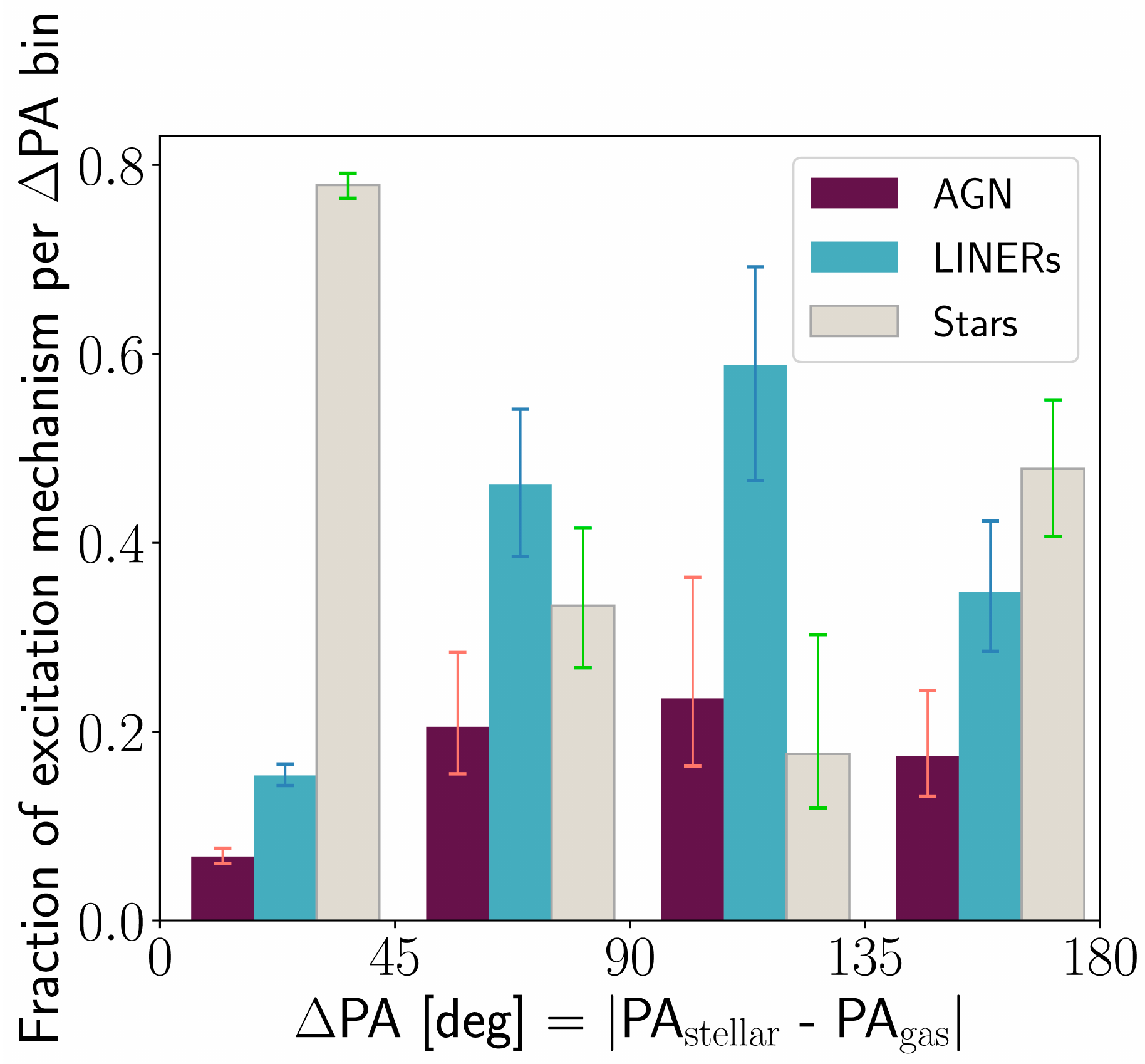}
	\caption{Fraction of galaxies with different excitation mechanisms divided per interval of misalignment angle ($\Delta$PA). Due to the low number of galaxies the histogram has been divided into $\Delta$PA $= 45^{\circ}$ bins which corresponds to a minimum of 17 total galaxies per bin. The higher $\Delta$PA bins tend to have a lower number of galaxies in them; using wide histogram bins ensures a minimum number of galaxies to achieve a reasonable statistical comparison. The histogram shows four bins: Bin1 = 0$^{\circ}\leq \Delta$PA $<$ 45$^{\circ}$, Bin2 = 45$^{\circ}\leq \Delta$PA $<$ 90$^{\circ}$, Bin3 = 90$^{\circ}\leq \Delta$PA $<$ 135$^{\circ}$ and Bin4 = 135$^{\circ}\leq \Delta$PA $\leq$ 180$^{\circ}$. Starforming galaxies tend to dominate the excitation mechanism in aligned galaxies while AGN and LINER excitation is relatively more common in misaligned galaxies ($\Delta$PA $\geq 45^{\circ}$). The error bars correspond to the 68\% confidence intervals using a beta distribution quantile technique \cite{cameron11}. The absolute values (in number of galaxies) for each histogram bar is given for AGN (A), LINERs (L) and Stars (S) as: Bin1 (A 67$^{+9}_{-7}$; L 152$^{+12}_{-11}$; S 770$^{+13}_{-14}$), Bin2 (A 8$^{+3}_{-2}$; L 18$^{+3}_{-3}$; S 13$^{+3}_{-3}$), Bin3 (A 4$^{+2}_{-1}$; L 10$^{+2}_{-2}$; S 3$^{+2}_{-1}$), Bin4 (A 8$^{+3}_{-2}$; L 16$^{+4}_{-3}$; S 22$^{+4}_{-3}$) with absolute sample sizes of: Bin1 $=$ 989, Bin2 $=$ 39, Bin3 $=$ 17, Bin4 = 46.}
	\label{Fig2_main}
\end{figure}

Fig.~\ref{Fig2_main} shows the distribution of excitation mechanisms as a function of the stellar-gas kinematic misalignment, for the subsample of \stellargasExcNum\ galaxies that have an excitation classification (BPT, broad lines or catalogue) and measured $\Delta$PA. We divided the sample into 45$^{\circ}$ bins corresponding to a minimum of 17 galaxies per bin. In each bin we measure the fraction of each of the three excitation mechanisms (AGN, star formation or LINERs). Adding the fractions for all mechanisms gives a total of 1 in each $\Delta$PA bin. The bin with the lowest $\Delta$PA values ($0^{\circ}\leq \Delta$PA $<$ 45$^{\circ}$) is clearly dominated by star-forming galaxies (non-AGN and non-LINER). This is also the bin that has a strong contribution by late-type galaxies (Fig.~\ref{Fig1_main}). The three bins with higher $\Delta$PA values in Fig.~\ref{Fig2_main} are consistent with having the same fraction of AGN excitation among them, within the 68\% confidence level. There is a clear difference between the relative fraction of star-forming vs (AGN+LINERs) between the lowest $\Delta$PA bin and the three higher $\Delta$PA bins, with AGN + LINERs being relatively more common in the higher $\Delta$PA bins when compared with star-forming (non-AGN) galaxies. The distribution for the bin with the most misaligned galaxies ($135^{\circ}\leq \Delta$PA $<$ 180$^{\circ}$) suggests that star-forming galaxies become relatively more common again at counter-rotation, as compared to AGN and LINERs. A version of this figure divided according to morphology can be found in the Supplementary Information.

\begin{figure}
	\centering
	\includegraphics[width=0.95\linewidth]{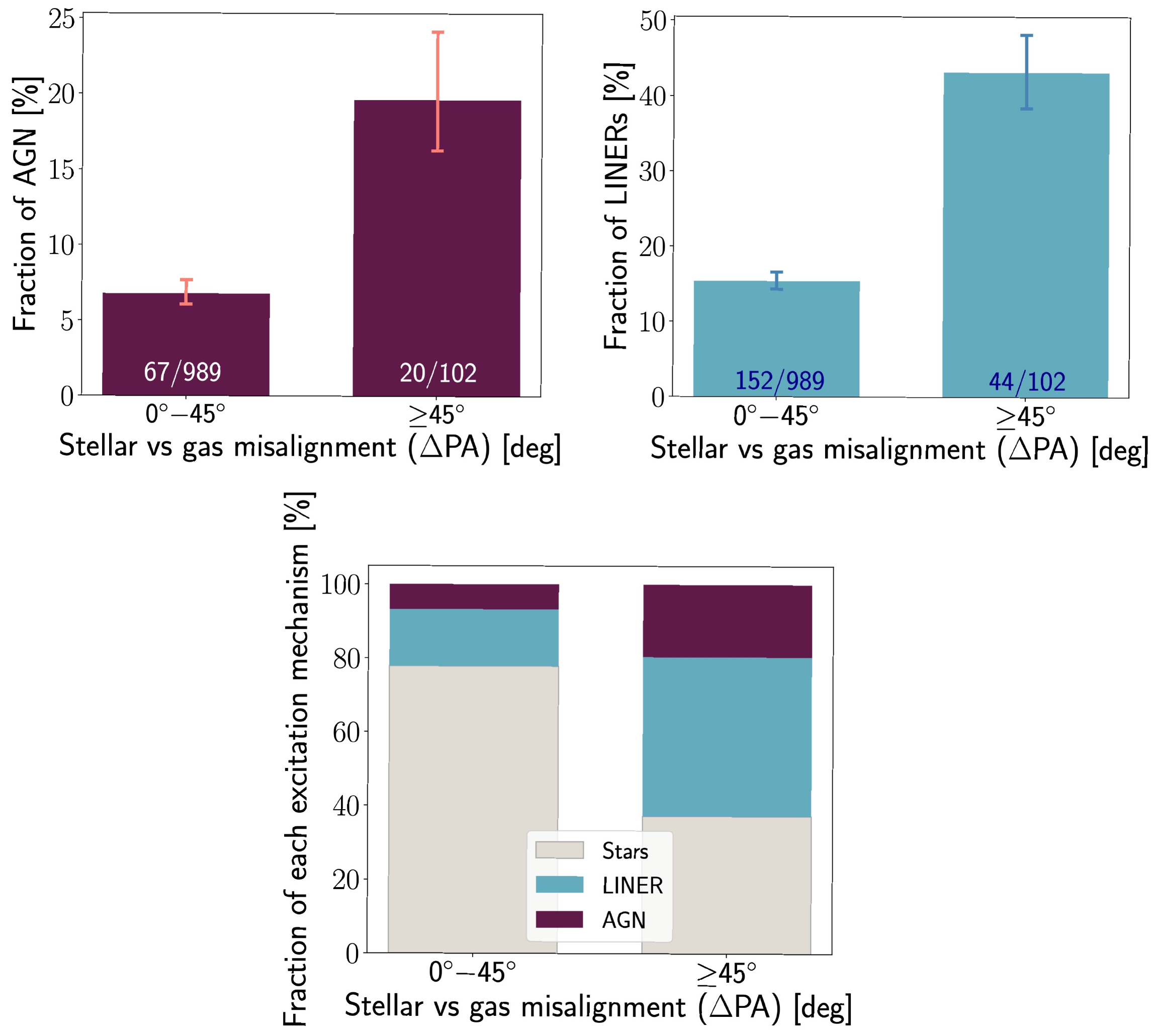}
	\caption{Fraction of AGN and LINERs in galaxies with aligned versus misaligned stellar to gas kinematics. The two top panels in the figure show the fraction of Active Galactic Nuclei (AGN - top left panel) and the fraction of Low Ionization Nuclear Emission Line Regions (LINERs - top right panel) in the sample of \stellargasExcNum\ galaxies from the SAMI survey \cite{croom21} with measured $\Delta$PA and excitation mechanism. The bottom panel shows the results for the entire sample divided into the three excitation mechanisms identified (AGN, LINER and star-forming galaxies). The galaxies are divided according to the angle difference between the stellar and ionised gas kinematic axis ($\Delta$PA = $\vert$PA$_{\rm stellar} - $PA$_{\rm gas}\vert$). Large kinematic misalignments between the stellar and gas rotation are a clear signature of an external accretion event, such as a galaxy merger or the accretion of gas from the immediate external environment of a galaxy. The two histogram bars show aligned ($0^{\circ}<\Delta$PA $< 45^{\circ}$) and misaligned ($45^{\circ} \leq \Delta$PA $\leq 180^{\circ}$) galaxies with error bars that indicate the 68$\%$ confidence intervals using a beta distribution quantile technique \cite{cameron11}. The numbers at the base of the histogram bars show the absolute number of AGN or LINERs compared with the total number of galaxies (sample size) in each bin. The absolute numbers for each bar in the order aligned/misaligned are: top left panel 67$^{+9}_{-7}$, 20$^{+5}_{-3}$; top right panel 152$^{+12}_{-11}$, 44$^{+5}_{-5}$, respectively. The figure shows that AGN occur at higher rates in galaxies with kinematic misalignment, with a significance of 99.7\% (3$\sigma$), showing that the presence of a stellar to gas kinematic misalignment is connected with a higher fraction of active supermassive black holes. LINERs show a similar trend to AGN. The increased fraction of LINER and AGN in misaligned galaxies contributes to a smaller fraction of star-forming galaxies in the misaligned group. 
	}
	\label{Fig3_main}
\end{figure}

To compare the AGN fraction in aligned versus misaligned galaxies, we use our sample of \stellargasExcNum\ galaxies. We classify these galaxies as aligned ($\Delta$PA $< 45^{\circ}$) or misaligned ($\Delta$PA $\geq 45^{\circ}$), and calculate the fraction of AGN in each of the groups. 
The result is shown in Fig.~\ref{Fig3_main}. The errors in the histogram of Fig.~\ref{Fig3_main} are the 68\% confidence intervals calculated using the beta distribution quantile technique for binomial population proportions, described in \cite{cameron11}. We use this more conservative approach (instead of the Poisson error estimate), as the Poisson error can underestimate the width of the confidence interval, in particular for small to medium size samples \cite{cameron11}. As can be seen in Fig.~\ref{Fig3_main} (top left panel), there is a higher fraction of AGN in misaligned galaxies (\AGNmisalignedForty) than in aligned galaxies (\AGNalignedForty). This difference is significant at the 99.7\% level. We reach similar conclusions when using an angle of $\Delta$PA = 30$^{\circ}$ as cutoff (Fig.~\ref{FigA2} of \nameref{sec:A1}). 

The top right panel of Fig.~\ref{Fig3_main} shows a similar analysis but for LINERs. Interestingly, galaxies with LINER excitation are also a higher fraction of misaligned galaxies (\LINERmisalignedForty) compared to \LINERalignedForty\ of aligned galaxies. The bottom panel shows the three excitation mechanisms together. Galaxies with young-star excitation show the opposite trend to AGN and LINERs -- the fraction of star formation excitation in misaligned galaxies (\STELLARmisalignedForty) is approximately half of the \STELLARalignedForty\ fraction  for aligned galaxies, due to the relatively larger proportions of LINER and AGN in misaligned galaxies. The differences between aligned and misaligned galaxies for LINERs and star-forming galaxies are significant at the 4.5$\sigma$ and 5$\sigma$ levels, respectively. The trend observed in all three panels is driven by galaxies in the field and in groups (i.e. non cluster environments). In the Supplementary Information we show for reference the results separated into field/group galaxies and cluster galaxies, respectively.

\section{Discussion}
\label{sec:discussion}
Our results show that misaligned galaxies are associated with a higher incidence of AGN and LINER excitation mechanisms. It is evident that the physical conditions in misaligned galaxies are favorable to AGN and LINERs. One likely reason is that a large stellar-gas misalignment is associated with external accretion, which often provides an additional gas supply to the galaxy (e.g. \cite{raimundo17}). Also, the presence of misaligned structures may facilitate the loss of gas angular momentum necessary to drive the gas to the nucleus \cite{thakar&ryden96}. Theoretical simulations have indeed predicted that strongly misaligned or counter-rotating structures promote gas inflow and the fuelling of supermassive black holes (e.g. \cite{negri14}, \cite{capelo&dotti17}, \cite{taylor18}). External gas accretion and its associated gas-stellar misalignment may therefore be able to provide both a supply of gas and the physical mechanim to drive the gas to the black hole and power AGN activity. The accretion of external gas may also promote shocks in the regions where the external gas interacts with the native material in the galaxy. LINERs may not reflect a single excitation mechanism, but could be  associated with either low-luminosity AGN, shocks or post-AGB stars (e.g. \cite{heckman80}, \cite{dopita&sutherland95}, \cite{binette94}). The fact that we see a higher fraction of LINERs in misaligned galaxies may be due to a large fraction of them being associated either with low-luminosity AGN or shocks after the external gas accretion, as seen in \cite{raimundo21}. 

The present work is the first time that the connection of external gas accretion and AGN activity has been observed in a large statistically significant sample of galaxies (Fig.~\ref{Fig3_main}). Note that the sample in Fig.~\ref{Fig3_main} only includes galaxies in which the PA$_{\rm stellar}$ and PA$_{\rm gas}$ can be measured. This, by definition, requires the presence of ionized gas in the galaxy. Our conclusion therefore is that for galaxies with ionised gas, the presence of a stellar to gas kinematic misalignment is connected with a higher fraction of AGN.

Aligned galaxies or counter-rotating galaxies appear to provide more favourable conditions for star-formation. Fig.~\ref{Fig2_main} shows that the two bins in which stellar excitation dominates are in the $\Delta$PA bins that contain aligned ($\Delta$PA $<45^{\circ}$) and counter-rotating ($145^{\circ}<\Delta$PA $\leq180^{\circ}$) galaxies. Due to dissipation, co-rotation and counter-rotation are also the two more stable gas dynamical configurations in galaxies with discs, after an external accretion event \cite{thakar&ryden96}. This may possibly be due to the fact that star formation often requires higher molecular gas densities which can more efficiently form in discs (e.g. \cite{bao22}) or to counter-rotating discs being more susceptible to instabilities \cite{thakar&ryden96}.  We note that due to AGN and LINER classification taking precedence in our selection (in that order), the galaxies that are classified as AGN may also have star forming activity and LINER activity as well, often outside the galaxy nucleus. Our classification of star-forming galaxies means that there is excitation by young stars only, but no detected AGN nor LINER activity. In fact, AGN are often accompanied by star formation \cite{rowan-robinson95}, and we observe that most of the AGN we identify (aligned or misaligned) have spaxels with detected LINER or star forming activity.

There is some further evidence from simulations that misalignments and AGN may be related. Work using the IllustrisTNG simulation, \cite{duckworth20} finds that low mass galaxies with misalignment have had a history of higher black hole luminosity and growth. However, \cite{duckworth20} highlights the difficulty in finding a correlation at low redshift. They find no difference between AGN and matched inactive galaxies at z$\sim$0 (consistent with observational results in \cite{ilha19}), possibly due to the different timescales of persistence for counter-rotation and black hole activity. They also predict that misaligned star-forming galaxies are more recent events than quiescent galaxies with active black holes, possibly because the latter had an earlier energy injection from the AGN that contributed to stopping star formation. If misaligned star-forming galaxies are indeed more recent, that would suggest that the lower fraction of star-formation only galaxies we observe in the misaligned galaxy group (Fig.~\ref{Fig3_main}, bottom panel) may not have had time to activate their black holes yet, but could become AGN in the future.

Studies of matched small samples of AGN and inactive galaxies have not found a significant difference between the gas-stellar misalignments for active and inactive galaxies. These results have been obtained when looking at the central kiloparsec of spiral galaxies \cite{dumas07}, with most of the AGN showing misalignments $< 20^{\circ}$, or when looking at samples from the MaNGA survey \cite{ilha19}. \cite{ilha19} find that the large scale gas kinematics is dominated by the host galaxy potential as opposed to the AGN. However, both \cite{dumas07} and \cite{ilha19} find hints of a more disturbed gas distribution for AGN than for inactive galaxies in the central 1 kpc, suggesting that these are the scales at which the AGN may affect the interstellar medium. Due to the spatial resolution of our data, we are not sensitive to the physical processes in the central hundreds of parsecs of our sample. Our work focuses on the large-scale gas kinematics which is dominated by the galaxy and its potential past interactions. We also ask a different question: whether the fraction of misaligned galaxies do or do not have a higher AGN fraction than aligned galaxies. Our finding of a higher fraction of AGN in misaligned galaxies does not necessarily mean that there will be a difference between populations when choosing a random sample of AGN and control galaxies. In our analysis, AGN are still more common (in numbers) in aligned galaxies, with \totalAGNalignedfrac\% of our AGN in galaxies with $\Delta$PA $<$ 45$^{\circ}$ (Fig.~\ref{FigA3}), increasing the chance of randomly selecting AGN in aligned galaxies. Additionally, the timescale for AGN activity flickering ($\sim 10^{5}$ yr, \cite{schawinski15}) is shorter than the lifetime of counter-rotating structures ($\sim$ Gyrs); this increases the chances of mismatching using a pair selection based on black hole activity. The approach in the present work uses as a first selection the physical feature (kinematic misalignment) that is longer-lived, as opposed to selecting based on AGN activity.

Morphology is an important factor when analysing stellar and gas alignment. Early-type galaxies (ellipticals and S0s) tend to have a broader distribution as a function of $\Delta$PA than late-type galaxies (spirals) which mostly have low $\Delta$PAs. 
Additionally, the majority of late-type galaxies are aligned, irrespective of the excitation mechanism (Fig.~\ref{FigA3} in \nameref{sec:A1}). This is somewhat expected based on the fact that external gas accretion (which creates strong misalignments) is thought to be less efficient in creating long-lived misaligned structures in late-type galaxies, due to the dissipation between the accreted gas and the higher native gas content of the galaxy (e.g. \cite{kannapan&fabricant01}, \cite{bassett17}). 

The number of galaxies with different excitation mechanisms and morphology are also shown in Table~\ref{Table2} of \nameref{sec:A1}. Note that these are galaxies for which the stellar and gas PA have been well determined in our analysis, which limits our sample to early-type galaxies with gas. This may be the reason why there are more AGN in early-type galaxies than in late-type galaxies in the sample, as opposed to what is commonly seen in X-ray selected AGN \cite{koss11}. Additionally our morphology division (see \nameref{sec:methods}) may also include some early-type spirals for which the distinction between S0 and early spiral is not clear in the SAMI classification. 
We see a trend as a function of morphology. We find that 75\% of misaligned galaxies are in early-type galaxies, in line with the predictions in simulations of a higher fraction of misalignment in early-type galaxies \cite{khim20}. AGN in misaligned hosts are also found at a higher percentage (95\%) in early-type galaxies (see Fig.~\ref{FigA4} in \nameref{sec:A1} and discussion in Section~\ref{sec:sub-samples} of \nameref{sec:methods}.

We have shown that for galaxies with gas, the presence of a stellar to gas kinematic misalignment is connected with a higher fraction of AGN at the 99.7\% (3$\sigma$) level of significance. A similar trend is seen for galaxies with LINER excitation. 
The fact that our sample consists of galaxies with gas suggests that the higher AGN fraction is not simply associated with the presence of gas but that there are other mechanisms connected with the formation and/or presence of misaligned structures that are linked with black hole fuelling. These mechanisms could be associated with the various stages of the process of external accretion, from early gas accretion, to gas transport in the galaxy and to the formation and long-term presence of misaligned structures. The results of our work suggest that the formation and presence of misaligned structures may be an important fuelling mechanism for black holes in early-type galaxies, which show a relative higher fraction of misalignment compared with late-type galaxies. This black hole fuelling mechanism is particularly relevant for all galaxies that undergo external interactions, especially at high redshift where minor and major mergers are expected to occur at a higher fraction.

\backmatter

\section{Methods}\label{sec:methods}

\subsection{Sample selection}
\label{sec:sample}

We selected our sample from the public data of the SAMI Galaxy Survey (\cite{croom12}, \cite{allen15}) with the Sydney-Australian-Astronomical-Observatory Multi-object Integral-Field Spectrograph. We used the public results from Data Release 3 (DR3, \cite{croom21}) that include observations of a total of 3068 unique galaxies. The SAMI instrument has a blue and a red arm covering the wavelength range of 3750 - 5750 and 6300 - 7400\AA\ with a velocity dispersion of 70.4 km s$^{-1}$ and 29.6 km s$^{-1}$ respectively (\cite{vandesande17}). The SAMI survey provides data cubes containing imaging and spectroscopy information on each of the galaxies. Each spatial pixel (`spaxel') in the cubes and maps is 0.5 $\times$ 0.5 arcsec with fibers distributed in a field-of-view of $\sim$15 arcsec diameter. The SAMI Survey has made many of their data products easily available, including the 3D cubes and higher level products such as stellar and ionised gas velocity maps and emission line maps. We retrieve the DR3 products available from the AAO Data Central \href{https://datacentral.org.au/}{https://datacentral.org.au/}. 
The sample of galaxies has redshifts in the range of 0.004 $<$ z $<$ 0.095. 

We use the stellar velocity maps available in DR3 \cite{vandesande17b}, that were obtained with the penalized Pixel Fitting method (pPXF) (\cite{cappellari&emsellem04}, \cite{cappellari17}). We use the maps of the Gaussian line-of-sight velocity distribution (i.e. with only two moments: velocity and velocity dispersion). In DR3 there are various available data cubes for each target which differ in the binning scheme used. We adopt the adaptively binned maps, which use the Voronoi method of \cite{cappellari&copin03} to bin the datacube spatially, to reach a median blue arm S/N$=$10. We find that this binning provides a good compromise between the spatial resolution and the S/N of the features we want to detect.
For the gas we use the ionised gas velocity maps obtained using the one-component fit, as we are interested in the overall large scale gas motions. The emission-line maps for DR3 have been obtained using LZIFU \cite{ho16}, a fitting code that subtracts the stellar continuum and simultaneously fits several emission lines in each spaxel of the data cubes. More details of how LZIFU is applied to SAMI can be found in (\cite{green18}, \cite{medling18}). In some cases the Voronoi binning as a function of  the continuum flux results in heavily binned cubes (with $\lesssim 5$ bins) which are not suitable for determining the gas kinematics. In those cases we also analyse the default (unbinned) gas flux data cubes. The SAMI team also provides the flux maps for several optical emission lines and the noise map in each spaxel, allowing us to use these maps and the S/N per spaxel to identify the spatially resolved excitation mechanisms. For the spatially resolved excitation maps we require a minimum S/N of 5 per spaxel to use an emission line.

We analyse all unique galaxies in the sample. For galaxies with more than one observation, we used the one with the highest quality as indicated from the flag `\textsc{isbest}' provided by the SAMI team. The DR3 also includes a morphological classification by different members of the team \cite{cortese16}, which we use here, where galaxies are divided into groups on a scale of integers and half-integers: (0=Elliptical; 0.5=Elliptical/S0; 1=S0; 1.5=S0/Early-spiral; 2=Early-spiral; 2.5=Early/Late spiral; 3=Late spiral; 5=? (unknown); -9=no agreement in the classification). 

\subsection{Kinematic angle analysis}
\label{sec:kin_analysis}
To determine the stellar and gas kinematic angles, needed to calculate the kinematic misalignment between gas and stars, we use the SAMI stellar and ionised gas velocity maps.

We calculate the global kinematic position angle (PA) for the stars and the gas using \textsc{fit$\_$kinematic$\_$pa} \cite{krajnovic06}. The SAMI team have calculated the PAs for the DR2 sample \cite{bryant19}. However, we carried out an independent analysis because we wanted to identify counter-rotation in particular, and therefore needed a consistent orientation for what we considered to be receding side and approaching side of the stellar and gas rotation. The global kinematic position angles are defined as the orientation of the mean motion of the stars or the gas as measured across the full spatial extent of the observed velocity maps. The kinematic PA can be understood as the angle between the North and the line that connects the absolute maxima of the velocity in the 2D velocity map (shown as the green line in Fig.~\ref{FigA1} in \nameref{sec:A1}). We define the PA to be measured from the North direction to the maximum velocity values in the approaching side of the rotation (corresponding to blueshifted velocities on the velocity maps). The PA values are measured from the full spatial size velocity map and, as defined in our work, vary between 0 - 360$^{\circ}$. The code \textsc{fit$\_$kinematic$\_$pa} calculates the PA by minimising the difference between the velocity maps and a bi(anti)symmetric version of the velocity map with respect to the zero velocity line (shown as a black dashed line in Fig.~\ref{FigA1} in \nameref{sec:A1}). More details on the code can be found in \cite{krajnovic06}. The PA difference between the stellar and gas rotation, $\Delta$PA, is determined from calculating $\Delta$PA = $\vert\Delta$PA$_{\rm stellar}$ - $\Delta$PA$_{\rm gas}\vert$ and varies from 0 - 180$^{\circ}$. With our approach we can detect the difference between co-rotation, ($\Delta$PA = 0$^{\circ}$), and counter-rotation ($\Delta$PA =180$^{\circ}$).  

\subsubsection{Stellar and gas kinematic position angle}
\label{sec:PAs}
We first applied a quality cut on the stellar velocity maps and only used spaxels where the uncertainty in the velocity V$_{\rm err} < 30$ km s$^{-1}$, the velocity dispersion is $\sigma > 35$ km s$^{-1}$ (corresponding to at least half of the instrumental FWHM), and the velocity dispersion uncertainty is $\sigma_{\rm err} < (\sigma \times 0.1 + 25)$ km s$^{-1}$, similar to the quality cuts of the SAMI team \cite{vandesande17b}. 
Due to the Voronoi binning method to increase the S/N, in some cases the data cubes are composed of a single (or a few) Voronoi bins. This makes it difficult to obtain a successful fit with \textsc{fit$\_$kinematic$\_$pa}. As a first cut, we require that the galaxy has at least 8 unique bins and 30 spaxels that obey the quality criteria above, for a PA fit to be carried out. We also automatically exclude fit results in which the uncertainty in the PA is $>30^{\circ}$. This cutoff was based on testing and visual confirmation. 
After the fit, we visually inspect all the maps to remove a small minority of cases in which the code is not able to accurately fit the velocity maps, for example when the best-fit PA indicated by the code is offset from the direction of rotation that is visually identified from the velocity maps. 

The method to find the kinematic PA of the gas is similar to that for the kinematic PA of the stars. We use the DR3 gas velocity maps determined for the binned and the unbinned data cubes, giving preference to the results for the binned data cubes unless the cubes are heavily binned (total bins $\lesssim 5$). We set a quality cut and only use spaxels where the velocity uncertainty V$_{\rm gas\_err} < 30$ km s$^{-1}$ and that have S/N $> 5$ in the gas flux measurement. Similarly to the stellar kinematic PA determination, we do a visual check of the final results. An example of the result of \textsc{fit$\_$kinematic$\_$pa} is shown in Fig.~\ref{FigA1} of \nameref{sec:A1}, illustrating a galaxy with stellar-to-gas counter-rotation ($\Delta$PA $\sim 180$ $^{\circ}$). 

\subsubsection{Fraction of misalignment}
\label{sec:misalignment}

To identify a difference in the stellar and gas kinematic angles ($\Delta$PA), we require all our galaxies to have determined PAs, i.e. to have been successfully fitted with \textsc{fit$\_$kinematic$\_$pa}. This results in \stellarPAnum\ unique galaxies for which a stellar kinematic PA can be determined and \gasPAnum\ unique galaxies for which a gas kinematic PA can be determined. The number of galaxies with both PA$_{\rm stellar}$ and PA$_{\rm gas}$ is \stellargasPAnum\ (Table~\ref{Table1} of \nameref{sec:A1}). The excluded galaxies for which PAs cannot be determined, tend to have a slight tendency to have lower photometric ellipticities (from Sersic fitting, \cite{owers19}) which could indicate lower inclinations (i.e. closer to face-on). The excluded galaxies are mostly late type spirals with lower stellar mass (M$_{*}$ $\sim$10$^{8}$ - 10$^{9.5}$ M$_{\odot}$) that fail the selection because both PA$_{\rm stellar}$ and PA$_{\rm gas}$ cannot be accurately determined, or early-type galaxies that fail the selection because PA$_{\rm gas}$ cannot be accurately determined (due to insufficient S/N in the gas emission lines).

Our findings can be directly compared with \cite{bryant19}, who used SAMI Data Release 2 (DR2) to do an analysis of the stellar vs gas kinematic misalignments. Their sample comprised a smaller number of galaxies (1213) compared to the sample used here (3068), however we can compare the percentage of misalignment between gas and stars. 

It is not obvious what angle to use to separate `aligned' from `misaligned' galaxies. As can be seen in Fig.~\ref{Fig1_main}, the $\Delta$PA distribution is close to continuous and there is no clear angle separation. Aligned galaxies may have a small angle misalignment due to small scale perturbations or uncertainties in defining the PA and may not always have $\Delta$PA $\sim 0$. Since our main goal is to study the properties of misaligned galaxies, we take a more conservative approach to separate the two populations and adopt an angle of $\Delta$PA = 45$^{\circ}$. This is a higher misalignment than the $\Delta$PA = 30$^{\circ}$ commonly used in the literature (e.g. \cite{davis11}),
because we want to be certain that the misalignment is real.

We find that out of the galaxies for which the kinematic position angles can be measured, \misalignedFracForty\ of them show a misalignment $\geq 45^{\circ}$ and \misalignedFracThirty\ of them show a misalignment $\geq 30^{\circ}$, which is slightly higher but marginally consistent within the errors with the 11$\pm1$\% of galaxies with $\geq 30^{\circ}$ misalignment found by \cite{bryant19}. 

To determine the effect of morphology we divide the sample into early-type and late-type galaxies using the SAMI classification criteria: 0 $<=$ early type $<=$ 1.5 and late type $>$ 1.5 to maximise the number of galaxies in our sample. We find a fraction of misalignment ($\Delta$PA $\geq 45^{\circ}$) of \misalignedFracEarlyFortyMyCut\ for early-type and \misalignedFracLateFortyMyCut\ for late-type galaxies. The distribution of $\Delta$PA for early- and late-type galaxies can be seen in Fig.~\ref{Fig1_main}. For comparison purposes, we also calculate the fraction of misaligned galaxies using a cutoff angle of 30$^{\circ}$. We find a fraction of misalignment ($\Delta$PA $\geq 30^{\circ}$) of \misalignedFracEarlyThirtyMyCut\ for early-type and \misalignedFracLateThirtyMyCut\ for late-type galaxies. \cite{bryant19} used a slightly different criterion for the identification of early type galaxies: (0 $<=$ early type $<$ 1.5 and late type $>$ 1.5) to have a clear separation between early-types and late-types but at the cost of providing a smaller sample. Using the same morphology criteria as \cite{bryant19} we find a similar fraction of late-type galaxies with misalignment (\misalignedFracLateThirtyBryant), compared with 5$\pm$1\% by \cite{bryant19}, and a slightly lower fraction of misaligned early-types (\misalignedFracEarlyThirtyBryant) compared with 45$\pm$6\% by \cite{bryant19}, but still consistent within the 68\% confidence intervals.  

Our sample of galaxies with measured $\Delta$PA is \earlyPAfrac\ early-type galaxies and \latePAfrac\ late-type galaxies, with the remaining 2\% being unclassified galaxies. If one considers the final sample of galaxies with measured $\Delta$PA and excitation (Table~\ref{Table2} of \nameref{sec:A1}), the fractions are somewhat different, with \earlyPAExcfrac\ of early-type galaxies and \latePAExcfrac\ of late-type galaxies. The relative increase in the fraction of late-type galaxies compared with early-type galaxies in the final sample is due to the requirement of measuring the line ratios for the BPT diagrams at S/N$> 5$, which is favoured in environments with more gas and more star-formation, such as in spiral galaxies. In fact, out of the 220 galaxies with PA$_{\rm stellar}$ and PA$_{\rm gas}$ but no BPT classification, 68\% are early-type galaxies. These 220 galaxies fail the BPT classification in 99\% of the cases due to not having enough spaxels with S/N $>5$ in either the H$\beta$ or [O III] emission lines. The remaining 1\% are cases where there is not a clear and unique BPT classification. 

\subsection{Methods to identify AGN}
We use the following methods to identify AGN in the sample:\\
\indent$\bullet$ Spatially resolved Baldwin, Phillips \& Terlevich (BPT) diagrams (\cite{baldwin81}, \cite{veilleux&osterbrock87});\\
\indent$\bullet$ Presence of broad emission lines in the nuclear spectrum;\\
\indent$\bullet$ Cross-matching with known AGN catalogues. 

\noindent We discuss each of these different methods below.

\subsubsection{BPT diagram}
\label{sec:BPT}
The SAMI DR3 includes the total fluxes (in each spaxel) of several optical emission lines, e.g.: [O II]$\lambda\lambda$(3727, 3729), H$\beta$, [O III]$\lambda$5007, [N II]$\lambda$6583, H$\alpha$, [S II]$\lambda$6716, and [S II]$\lambda$6731. The SAMI team uses LZIFU \cite{ho16} to fit the emission lines and LZCOMP \cite{hampton17} to determine the optimal number of components for each one of them. In our analysis we use the emission-line maps corresponding to the `recommended' number of components of the multi-Gaussian fits, according to the SAMI team. These maps include the total flux emitted in each line (summed over all the components), which is what we use to create our emission-line diagnostics. We choose the unbinned data cubes labelled as `default' to have more sensitivity to the spatial variations of excitation across the field of view.

The excitation diagnostics we use are based on the ratio between the flux of several emission lines (\cite{baldwin81}, \cite{veilleux&osterbrock87}), also known as BPT diagrams. We use several line ratios: [O III]$\lambda$5007/H$\beta$ vs [NII]$\lambda$6583/H$\alpha$, [O III]$\lambda$5007/H$\beta$ vs [SII]($\lambda$6716 + $\lambda$6731)/H$\alpha$ and [O III]$\lambda$5007/H$\beta$ vs [OI]$\lambda$6300/H$\alpha$. To identify excitation mechanisms based on these line ratios, we use the theoretical regions defined by \cite{kewley06}, that can be used to separate AGN excitation from excitation associated with star-formation, for example. These diagrams and theoretical regions are shown in Fig.~\ref{FigA5} of \nameref{sec:A1}. We use the diagrams to separate spatial regions in our galaxies with gas excitation by young stars (H II regions), AGN or LINERs (Low-Ionization Nuclear Emission Line Regions).

Our goal is to identify AGN. However, as we are dealing with data cubes, each spaxel or bin defines one data point in the BPT diagram. For example, in galaxies with weak AGN, the spaxels in the nuclear region may be consistent with AGN excitation, while the galaxy outskirts may be dominated by star forming regions.  This trend is often evident in spatially resolved maps of BPT line ratio diagnostics in nearby AGN seen by \cite{xia18}.
For those ``composite" cases we still want to identify the galaxy as `AGN' as it meets our criteria for the presence of a currently active supermassive black hole. We do the same for LINER excitation, i.e. if a LINER-like region is present in the galaxy we use the classification label `LINER', even if this region might not necessarily be powered by an accreting black hole. Some of these LINERs may be low-luminosity AGN (e.g. \cite{ho97}), but others may have line contributions from shocks (e.g. \cite{heckman80}, \cite{dopita&sutherland95}) or from stars in the post-AGB phase (e.g. \cite{binette94}). We keep AGN and LINERs as two separate classes.

For our analysis we first exclude spaxels from each line emission map that have S/N $< 5$ in their line flux, and only consider maps with 5 or more valid spaxels for the automatic classification. To be classified, the galaxy needs to have a consistent classification in the [O III]/H$\beta$ vs [NII]/H$\alpha$ diagram plus at least one of the other two diagrams. We then use an automatic classification based on the percentage of pixels in each of the BPT regions, a similar approach to that used by \cite{wylezalek18} for the MaNGA survey. To take into account a possible AGN contribution to the excitation in the `composite' region, we use a similar `AGN weight' parameter and weight division as \cite{wylezalek18}. We attribute a weight of 80\% to AGN regions and 20\% to composite regions in the [O III]/H$\beta$ vs [NII]/H$\alpha$ diagram, as composite regions may have a contribution from AGN excitation in addition to  star-formation (e.g. \cite{kewley06}). We define the AGN weight to be AGN$_{w} =$ 0.2$\times$f$_{\rm [N II]\_composite}$ + 0.8$\times$f$_{\rm [N II]\_AGN}$, with f$_{\rm [N II]\_composite}$ being the fraction of pixels in the composite region and f$_{\rm [N II]\_AGN}$ being the fraction of pixels in the AGN region of the [NII]/H$\alpha$ diagram (see Fig.~\ref{FigA5} of \nameref{sec:A1}). 

To find the best cutoffs for the classification we ran several tests followed by visual inspection and defined the following criteria for classification into AGN, LINER and star forming galaxy, which are somewhat different from those in \cite{wylezalek18}. The variables below refer to fractions of spaxels in each region of the various BPT diagrams, as illustrated in Fig.~\ref{FigA5} of \nameref{sec:A1}: \\
$\bullet$ AGN: AGN$_{w} > 0.05$ and (f$_{\rm [S II]\_AGN} > 0.05$ or f$_{\rm [O I]\_AGN} > 0.25$)\\
$\bullet$ LINER: AGN$_{w} > 0.05$ and [(f$_{\rm [O I]\_AGN}$ + f$_{\rm\_[O I] LINER}) > 0.25$ or (f$_{\rm [S II]AGN} $ + f$_{\rm [S II]\_LINER}) > 0.05$] and a non-AGN classification according to the criteria above. \\
$\bullet$ Star forming: (f$_{\rm [N II]\_SF}$ +  f$_{\rm [N II]\_comp}) > 0.05$ and a Non-AGN and Non-LINER classification according to the criteria above. \\

These criteria are applied to maps with 50 or more valid spaxels. For less than 50 spaxels, we define more stringent criteria where instead of percentages of 5\% in the criteria above, we require percentages of 25\%.
For all the maps where an automatic classification is not reached, we inspect the nuclear spectra for each galaxy manually and decide on a classification on a case by case basis. For the manual classification we inspect the nuclear 1D spectra (integrated within an aperture of 1.4 arcsec) provided for each galaxy as part of DR3. This inspection is particularly important for higher redshift galaxies where the [S II] emission line doublet may be outside the wavelength range covered. In those cases, a visual inspection is important since one of the BPT diagrams will not have data points for an automatic classification. 

In Fig.~\ref{FigA6} of \nameref{sec:A1} we show an example of the diagnostics we use, and select three different galaxies to highlight the different excitation diagnostics and classifications.

Out of the \BPTnum\ galaxies with a BPT classification, \BPTAGN\ of them are AGN (\BPTAGNFrac\%), \BPTLINER\ are LINERs (\BPTLINERFrac\%), and the remaining \BPTSTAR\ are star-forming galaxies (\BPTSTARFrac\%) with no clear signatures of either AGN or LINER emission. 
The fraction of AGN we find is lower than that found for the MANGA survey (11$\%$) \cite{wylezalek18}. \cite{wylezalek18} classifies some targets as AGN that we here would classify as LINER based on the [S II]/H$\alpha$ line ratio. This is because \cite{wylezalek18} uses an additional diagnostic to distinguish these LINERs from AGN. We emphasize that we used a conservative classification with the goal of identifying highly likely AGN. It is possible that a fraction of LINERs or of the star-forming galaxies in our classification may have contributions beyond young stars and could have some contribution from AGN excitation. We show some examples of this in Section~\ref{sec:broad_lines}. 

If we consider only these galaxies in which the BPT diagram can be measured, we obtain a  fraction of misaligned galaxies ($\Delta$PA $\geq 45^{\circ}$) of \misalignedBPTForty\, which is similar to the \misalignedFracForty\ that was determined in Section~\ref{sec:misalignment}. This shows that selecting galaxies with BPT information does not bias against the fraction of misaligned galaxies.

\subsubsection{Broad emission lines}
\label{sec:broad_lines}
Broad emission lines arise in the high velocity gas very near the accreting supermassive black hole and are a strong confirmation of the presence of an AGN in the galaxy. The spectral region of the SAMI observations cover both the H$\alpha$ $\lambda$6563\,\AA\ and H$\beta$ $\lambda$4861\,\AA\ emission lines, which are broad (FWHM $> 1200$ km/s) in type 1 AGN (e.g. \cite{hao15}). One of the SAMI survey data products is the 1D spectrum of the nuclear fibre (within an aperture of 1.4 arcsec) for each of the target galaxies. We use these spectra to search for the presence of broad H$\alpha$ which is expected to be the strongest broad emission line in the covered wavelength range. While this method only detects type 1 AGN, 
the detection of a broad component is a strong confirmation of the presence of an AGN in the galaxy.

We fit all the 1D spectra using BADASS
(\cite{sexton21}), a Bayesian fitting tool that fits multi-components (e.g. host galaxy stellar light, narrow and broad emission lines) to observed galaxy spectra \href{https://github.com/remingtonsexton/BADASS3}{https://github.com/remingtonsexton/BADASS3}. We use a conservative selection criterion and consider that galaxies have an AGN when a broad line (FWHM $> 1200$ km/s) is detected in the spectral fit and then confirmed visually.  Cases where a weak broad component is detected but not clearly identified visually are excluded and not classified as AGN to avoid false positives. An example of a galaxy with a broad H$\alpha$ line is shown in the Supplementary Information.

We find a total of \blAGN\ broad-line AGN. Out of the \blAGN, \blAGNBPTAGN\ have been identified as AGN from the BPT analysis, while \blAGNBPTSF\ were classified as star-forming and \blAGNBPTLINER\ as LINERs from the BPT diagram analysis. The misclassification may be due to the degeneracy in obtaining accurate fluxes for the narrow emission lines with LZIFU when a broad line is present. It may also highlight the fact that some star-forming or LINER excitation from the BPT diagram may have a small nuclear AGN contribution, as mentioned in the previous section, or that some Seyfert 1 galaxies may have very weak narrow line regions. Note that we give precedence to AGN classification as that is the focus of this work. For example, a galaxy that is classified as star-forming in the BPT diagram but that has a broad emission line will be classified as an AGN. 

\subsubsection{Cross matching with AGN catalogues}
To further identify AGN, we cross-matched the SAMI DR3 catalogue with several all-sky AGN catalogues in the X-rays, far-infrared and radio wavelengths which are known to be less biased methods of identifying AGN, especially if the nucleus is obscured. We used two X-ray AGN catalogues: the Second ROSAT all-sky survey (2RXS) \href{http://vizier.u-strasbg.fr/viz-bin/VizieR?-source=J/A+A/588/a103}{http://vizier.u-strasbg.fr/viz-bin/VizieR?-source=J/A+A/588/a103} \cite{boller16} and the Swift/BAT 70 month AGN X-ray catalogue \href{http://vizier.u-strasbg.fr/viz-bin/VizieR?-source=J/ApJS/233/17}{http://vizier.u-strasbg.fr/viz-bin/VizieR?-source=J/ApJS/233/17} \cite{ricci17}. We also used two WISE-based far-infrared AGN catalogues:  the WISE AGN catalog (90\% confidence level) based on the AllWISE catalogue \cite{assef18} \href{http://vizier.u-strasbg.fr/viz-bin/VizieR?-source=J/ApJS/234/23}{http://vizier.u-strasbg.fr/viz-bin/VizieR?-source=J/ApJS/234/23} and the AGN identified as `highly likely' from a combination of WISE, ROSAT and 2MASS data in the catalogue of \cite{edelson&malkan12} \href{http://vizier.u-strasbg.fr/viz-bin/VizieR?-source=J/ApJ/751/52
	}{http://vizier.u-strasbg.fr/viz-bin/VizieR?-source=J/ApJ/751/52
}. We also used radio-loud AGN identified from the catalogue of \cite{best&heckman12} \href{http://vizier.u-strasbg.fr/viz-bin/VizieR?-source=J/MNRAS/421/1569}{http://vizier.u-strasbg.fr/viz-bin/VizieR?-source=J/MNRAS/421/1569}. 
We used \textsc{topcat} \cite{taylor05} to cross match our sample with all the above catalogues using a matching radius of 7.5 arcsec corresponding to half of the field of view of the SAMI datacubes. We identify 23 AGN in the SAMI sample. None of these 23 AGN had been classified as AGN based on the BPT diagram, and only 2 of them had been identified from the presence of broad emission lines. This highlights the importance of using multi-wavelength data in the identification of AGN.

\subsection{Properties of the sub-samples}
\label{sec:sub-samples}
We investigate how aligned and misaligned AGN and galaxies are distributed as a function of morphology in Fig.~\ref{FigA4} of \nameref{sec:A1}. A larger percentage (58\%) of our AGN are in early-type hosts, as opposed to 42\% in late-type hosts (Table~\ref{Table2} in \nameref{sec:A1}). However, if one looks at only the sub-sample of AGN found in aligned galaxies (bottom left panel of Fig.~\ref{FigA4} of \nameref{sec:A1}), the AGN do not show a strong preference between early-type and late-type galaxies. This suggests that once stars and gas are aligned, the AGN do not show a clear distinction between early-types and late-types, even though their host galaxies do. We note that external gas accretion will not always result in misalignment but in some cases will result in aligned stellar to gas configurations. A significant percentage of aligned early type galaxies in our sample may have had an external accretion event as well, as also seen by \cite{davis11}.

In this section we also show the distribution of gas velocity dispersion and stellar mass for the different sub-samples analysed in this work. We show that the trend we observe in the aligned vs misaligned AGN samples (Fig.~\ref{Fig3_main}) is not caused by the presence of AGN-driven outflows or by a stellar mass bias. Our results are discussed below.

Fig.~\ref{FigA7} of \nameref{sec:A1} shows the uniform spatially-averaged gas velocity dispersion ($\overline{\sigma_{\rm gas}}$) for each galaxy, divided into four panels, one for each sub-sample. If AGN-driven outflows were responsible for the large scale stellar to gas misalignment observed in the `misaligned' sample, we would expect to see a significant difference between the distribution of velocity dispersion, with misaligned AGN showing significantly higher velocity dispersion due to outflows. Since our analysis uses the large scale gas dynamics, we exclude the central 3{$''$} of each galaxy, corresponding to the upper limit on the SAMI DR3 Point Spread Function, to avoid possible unresolved nuclear AGN outflows. The velocity dispersion maps we use are the same as for the analysis of the gas PA, and therefore have the same quality cuts as those defined in Section \ref{sec:PAs} of \nameref{sec:A1}. As can be seen in Fig.~\ref{FigA7} of \nameref{sec:A1}, there is no significant difference between the distribution of average velocity dispersion for the AGN subsamples (aligned vs misaligned) nor between the misaligned AGN and misaligned non-AGN galaxy samples. To do a more quantitative comparison we used a Mann–Whitney U statistical test, to test if two samples are likely from the same underlying parent distribution. This test was used due to the relatively small size of our sample of misaligned AGN. When comparing each pair of samples, we obtain a p value of 0.1 for the comparison between aligned and misaligned AGN and a p value of 0.1 between misaligned AGN and misaligned non-AGN galaxies. With the null hypothesis that the distribution underlying both samples is the same, these p-values measure the evidence against the null hypothesis and indicate that the null hypothesis cannot be rejected. In other words, the sub-samples are consistent with being drawn from the same distribution and there is no evidence from this analysis that AGN outflows are causing the gas to be misaligned in the population of misaligned AGN. Instead, the misaligned gas is consistent with being the result of an external accretion event, as described in the previous sections. Interestingly, the sub-samples of aligned and misaligned non-AGN galaxies (right panels in Fig.~\ref{FigA7} of \nameref{sec:A1}) show a p-value of $\sim$10$^{-13}$, indicating that the null hypothesis (that the samples are drawn from the same distribution) can be rejected at a minimum confidence level of 95\%. This finding supports a different set of properties and/or evolution path between these two sub-samples of galaxies, such as a higher incidence of external accretion events in the misaligned galaxy sample, for example. A similar effect of rejecting the null hypothesis is seen between AGN (both in aligned or misaligned hosts) and non-AGN aligned galaxies.

We carry out a similar analysis for the stellar mass distribution of the different sub-samples, to evaluate if a trend of AGN with host galaxy stellar mass could cause the difference in aligned/misaligned AGN fractions that we observe. In Fig.~\ref{FigA8} of \nameref{sec:A1} we show histograms of the stellar mass distribution for each of the sub-samples: aligned AGN, misaligned AGN, aligned (non-AGN) galaxies and misaligned (non-AGN) galaxies. There is no significant difference between the stellar mass distribution for misaligned AGN as compared to both aligned AGN and misaligned galaxies. This indicates that the trend of a higher AGN fraction in misaligned galaxies is not driven by a stellar mass bias (for example due to having a higher number of AGN in more massive galaxies). Our Mann–Whitney U statistical tests show p-values of 0.5 and 0.2 for the AGN aligned/misaligned sub-samples and misaligned AGN/galaxy subsamples, respectively. This indicates that the samples are consistent with being draw from the same underlying distribution. Once again we find a p value = 0.008 for the comparison between aligned and misaligned non-AGN galaxies, which suggests that these are two distinct distributions at the 95\% confidence level (since the p value $<$ 0.05).\\

\bmhead{Data Availability}
All correspondence and requests for materials should be addressed to Sandra I. Raimundo (s.raimundo@soton.ac.uk). The data used in this study are available in the Australian Astronomical Observatory (AAO) Data Central repository: \href{https://docs.datacentral.org.au/sami/}{https://docs.datacentral.org.au/sami/}. Access to additional data can be done via a persistent repository: \href{https://erda.ku.dk/archives/dcf9b1543592f8fbd824cf1eeb733b4e/published-archive.html}{https://erda.ku.dk/archives/dcf9b1543592f8fbd824cf1eeb733b4e/published-archive.html}

\bmhead{Acknowledgments}

The authors would like to thank the referees for their constructive comments. This project has received funding from the European Union$'$s Horizon 2020 research and innovation programme under the Marie Sklodowska-Curie grant agreement No 891744 (S.I.R). This research has been financially supported by the Independent Research Fund Denmark via grant number DFF 8021-00130 (M.V). This paper includes data that has been provided by AAO Data Central (datacentral.org.au). The SAMI Galaxy Survey is based on observations made at the Anglo-Australian Telescope. The Sydney-AAO Multi-object Integral field spectrograph (SAMI) was developed jointly by the University of Sydney and the Australian Astronomical Observatory. The SAMI input catalogue is based on data taken from the Sloan Digital Sky Survey, the GAMA Survey, and the VST ATLAS Survey. The SAMI Galaxy Survey is funded by the Australian Research Council Centre of Excellence for All-sky Astrophysics (CAASTRO), through project number CE110001020, and other participating institutions. 
This research has made use of the VizieR catalogue access tool, CDS, Strasbourg, France (DOI : 10.26093/cds/vizier). The original description of the VizieR service was published in \cite{ochsenbein00}.
This research made use of Astropy, \href{http://www.astropy.org}{http://www.astropy.org} a community-developed core Python package for Astronomy \cite{astropy:2013}. 

\bmhead{Author Contributions Statement}
SIR conceived the study, carried out the analysis and wrote the paper. MM and MV wrote the paper, making an equal contribution to the paper. All authors discussed the results and their interpretation and commented on the manuscript at all stages.

\bmhead{Competing Interests Statement}
The authors declare no competing interests.


\clearpage

\section{Extended Data}
\label{sec:A1}

\begin{table}[b]
\centering
	    \includegraphics[width=8.0cm]{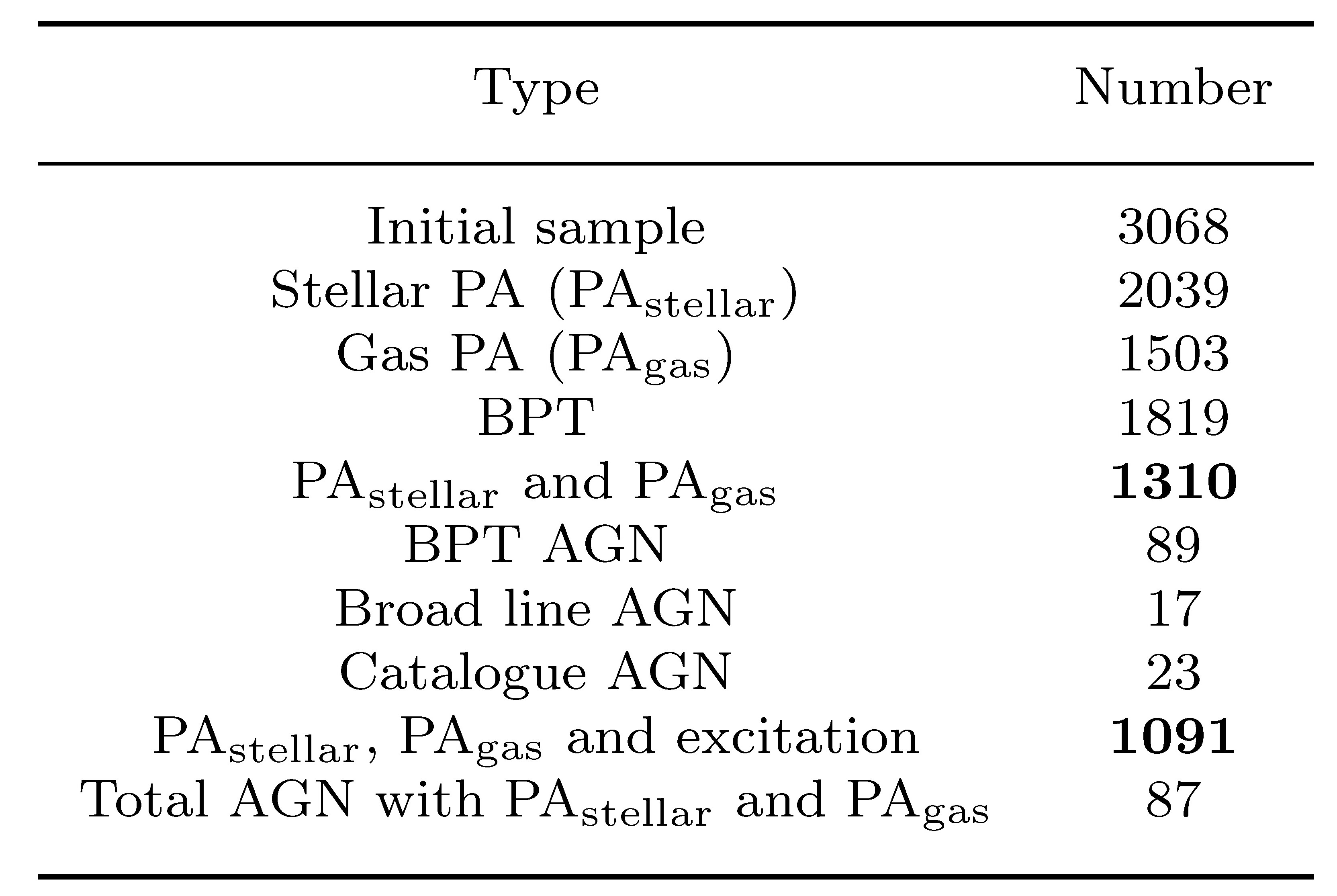}
\caption{Numbers for the different galaxy sub-samples used in this work. Note: there is some overlap in the AGN identification, as some AGN are identified both in the BPT diagram and with broad lines or with broad lines and catalogue matching. Text in bold highlights the two most important galaxy sub-samples.}
\label{Table1}
\end{table}

\begin{table}
	\centering
	\includegraphics[width=12.0cm]{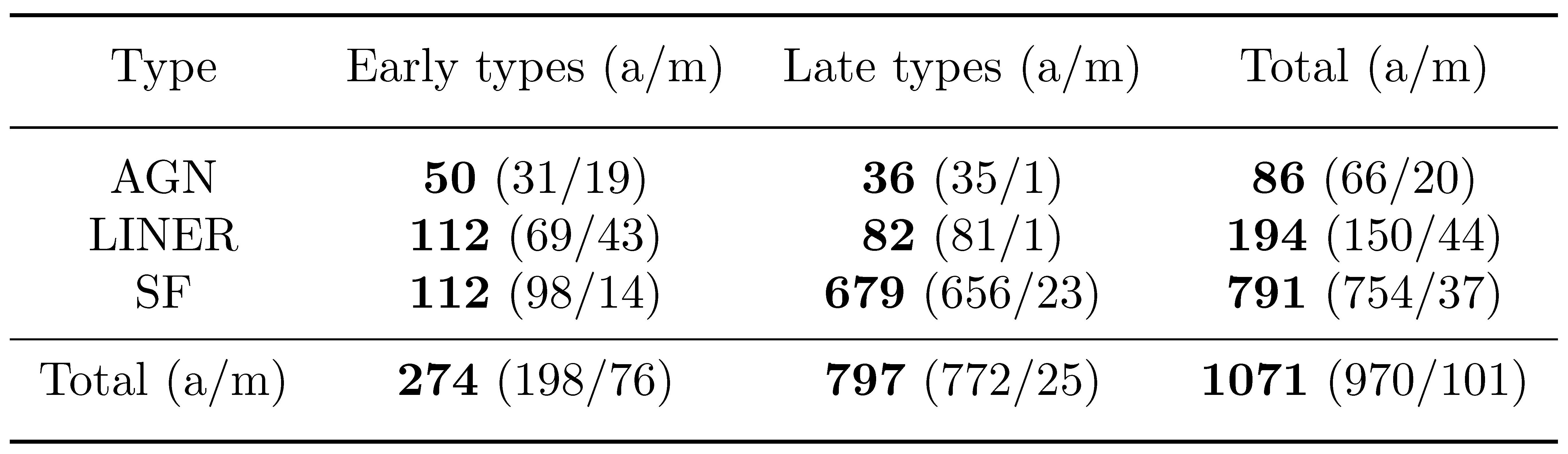}
\caption{Distribution of galaxies that have an excitation classification and PA$_{\rm stellar}$ and  PA$_{\rm gas}$ well constrained, divided according to morphology. `a/m' stands for aligned/misaligned numbers in each group of galaxies and `SF' for star-forming galaxies. Numbers in bold show the total number of galaxies in each classification group. The total number of galaxies is different from Table~\ref{Table1} because some of the SAMI galaxies do not have a clear morphological classification.}
\label{Table2}
\end{table}

\renewcommand{\thefigure}{A\arabic{figure}}
\setcounter{figure}{0}   

\begin{figure*}
	\centering
	\includegraphics[width=12cm]{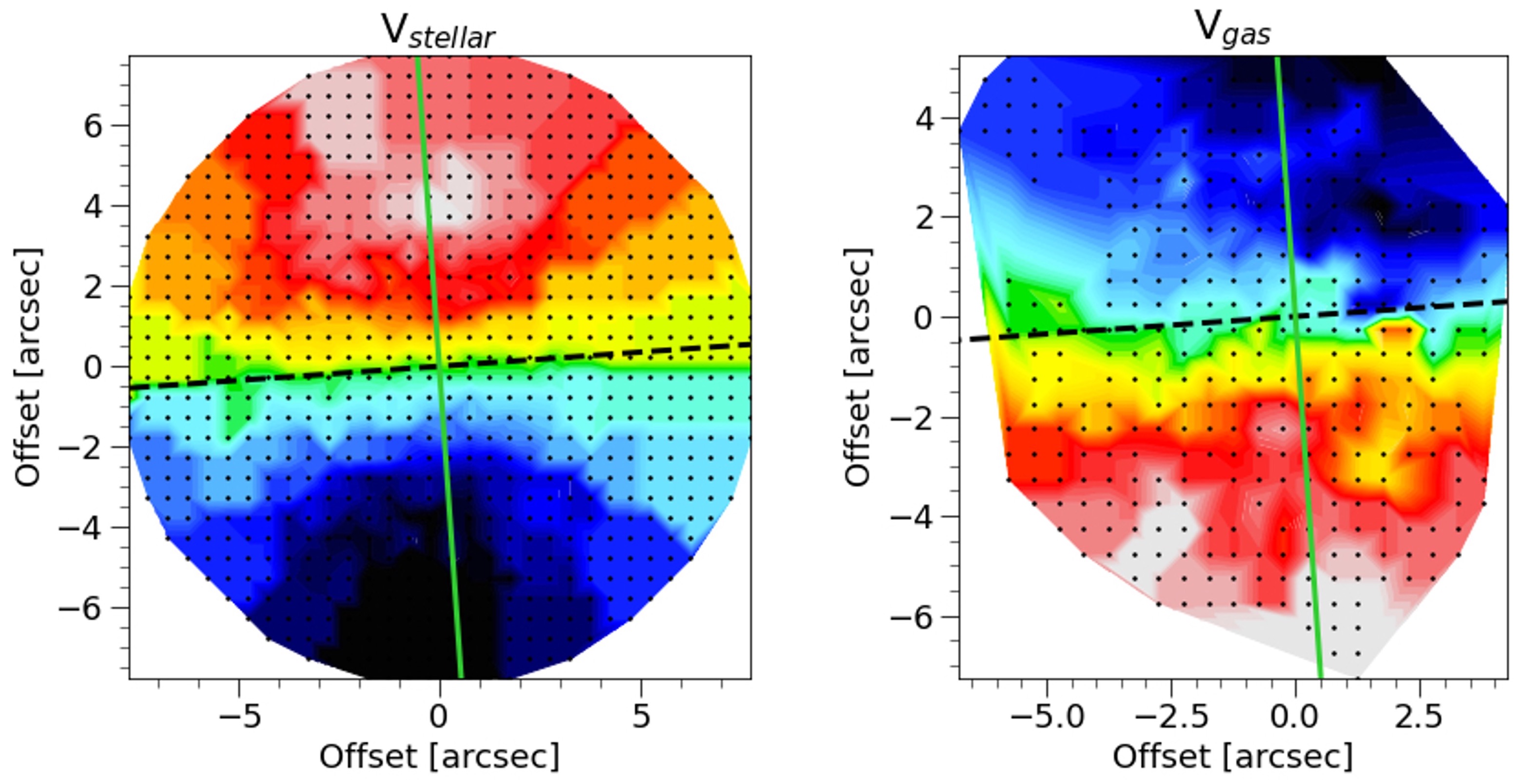}
	\caption{Result of the kinematic position angle (PA) determination for one of the galaxies in the sample, J145002.16+003443.6, or ID 93807 in the SAMI catalogue. Left: Map of stellar velocity. The best fit kinematic PA is shown by the green solid line. The value determined is PA$_{\rm stellar} = 184.0 \pm 1.5^{\circ}$ where the error corresponds to the 3$\sigma$ uncertainties of the fit. Right:  Map of gas velocity. The best fit kinematic PA is shown by the green solid line, PA$_{\rm gas} = 4 \pm 4^{\circ}$. The PA orientation is measured from North (PA = 0, up in the figure) to East and is per our definition oriented from the approaching (blueshifted) to the receding (redshifted) regions of the map. The dashed line shows the zero velocity line, which is the axis that \textsc{fit\_kinematic\_pa} uses to create the mirrored bi(anti)symmetric velocity map. The measured misalignment between the stellar and gas kinematic angles for this galaxy, $\Delta$PA = $\vert$PA$_{\rm stellar}$ - PA$_{\rm gas}\vert$ = 180$^{\circ}$, corresponds to counter-rotation of gas and stars.}
	\label{FigA1}
\end{figure*}

\begin{figure}
	\centering
	\includegraphics[width=1.0\linewidth]{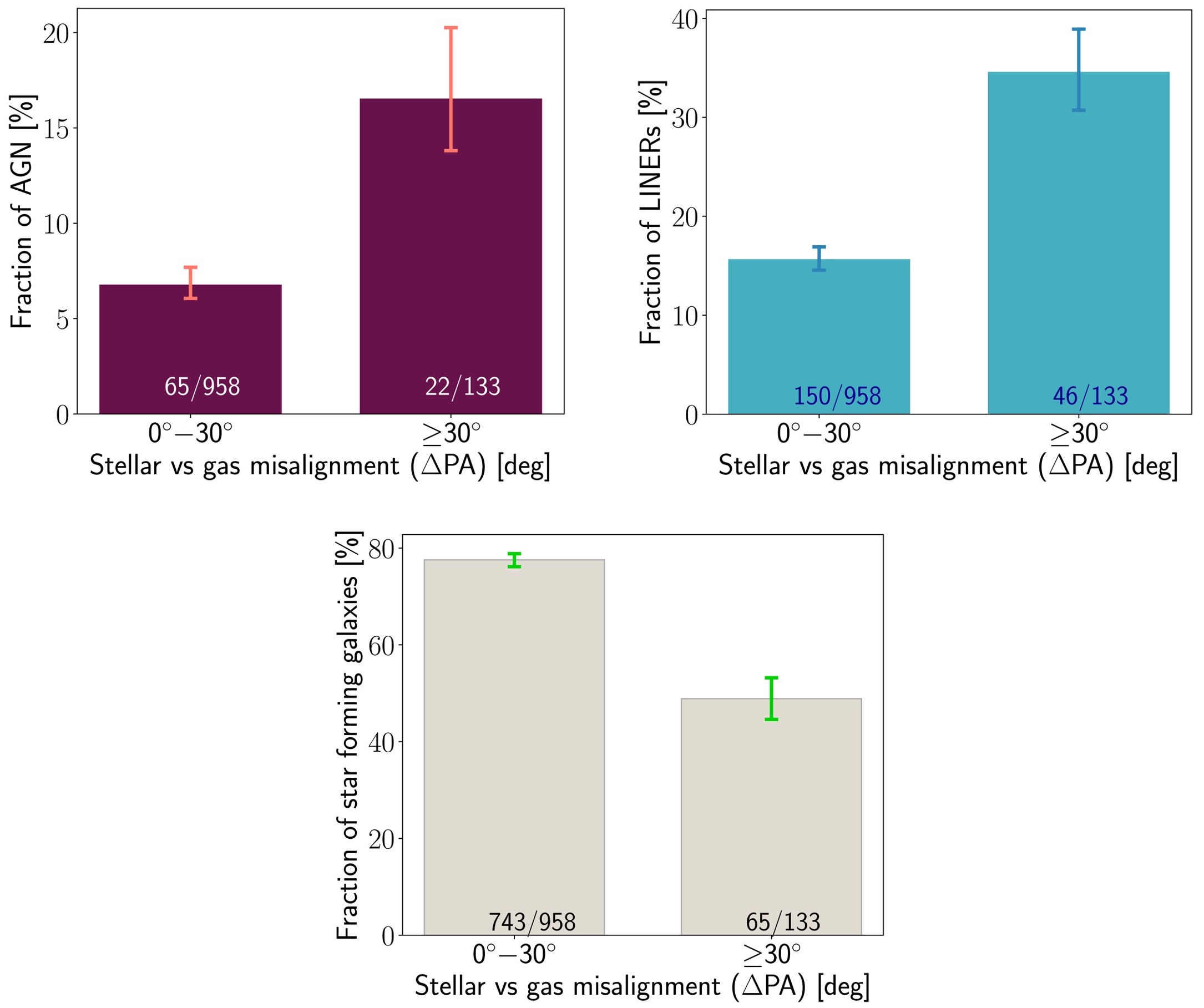}
	\caption{Similar to Fig.~\ref{Fig3_main} but using $\Delta$PA $= 30^{\circ}$ as the cut-off angle. The fraction of AGN is \AGNalignedThirty\ in aligned galaxies and \AGNmisalignedThirty\ in misaligned galaxies, the fraction of LINERs is \LINERalignedThirty\ in aligned galaxies and \LINERmisalignedThirty\ in misaligned galaxies and the fraction of star forming galaxies is \STELLARalignedThirty\ in aligned galaxies and \STELLARmisalignedThirty\ in misaligned galaxies. The absolute numbers for each bar in the order aligned/misaligned are: top left panel 65$^{+9}_{-7}$, 22$^{+5}_{-4}$; top right panel 150$^{+12}_{-11}$, 46$^{+6}_{-5}$; bottom panel 743$^{+12}_{-13}$, 65$^{+6}_{-6}$, respectively. The error bars correspond to the 68\% confidence intervals using a beta distribution quantile technique.}
	\label{FigA2}
\end{figure}

\begin{figure}[h]
	\centering
	\includegraphics[width=12cm]{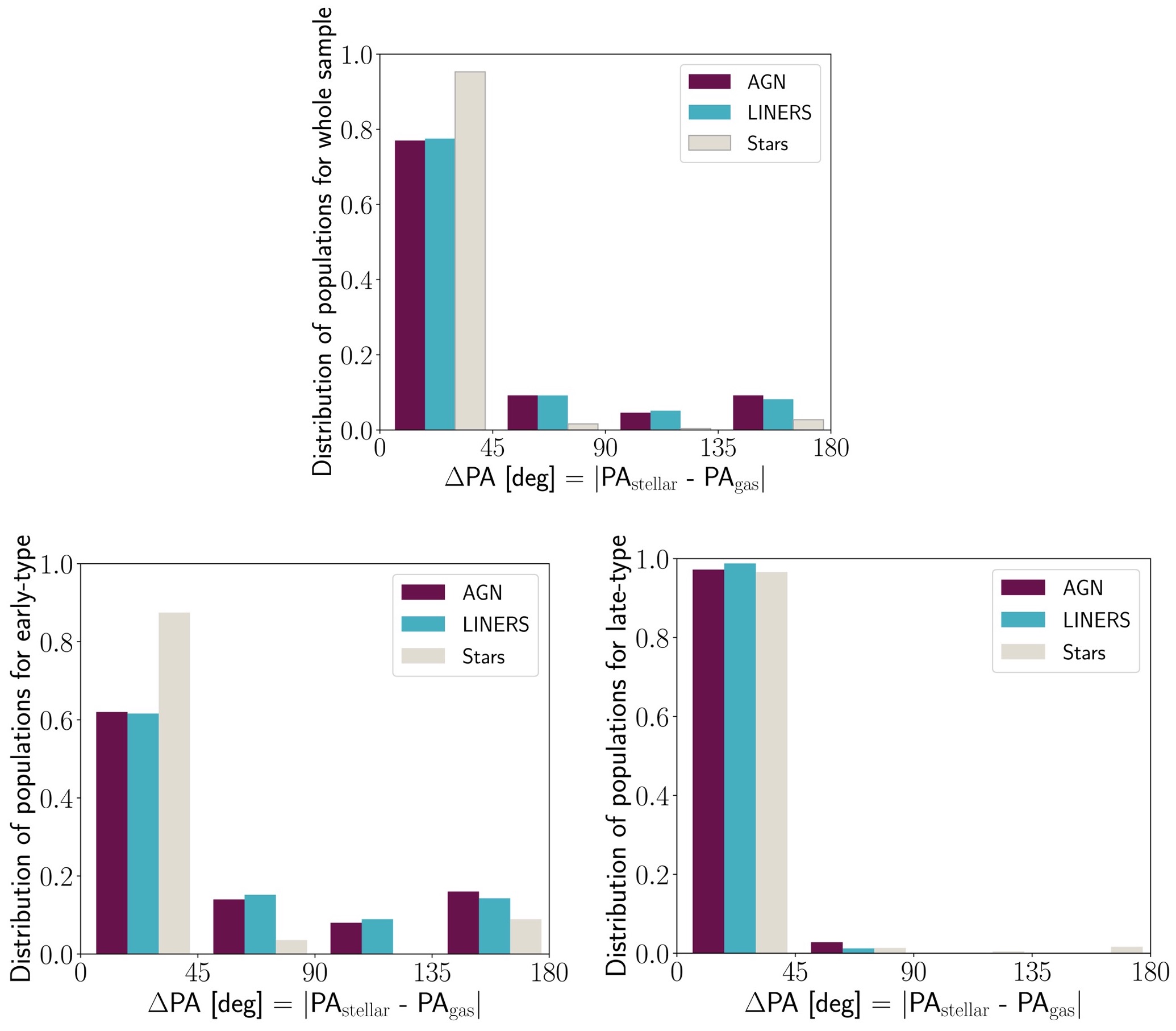}
	\caption{Fraction of galaxies with a specific excitation mechanism, as a function of the misalignment between the stellar and the gas kinematic angle ($\Delta$PA = $\vert$PA$_{\rm stellar}$ - PA$_{\rm gas}\vert$) and the galaxy morphology. The histogram bars of each colour add up to 1 across the $\Delta$PA distribution. The top panel shows the total population, the bottom left panel shows the early type galaxies and the bottom right panel shows the late type galaxies in the sample. Each panel is colour coded as a function of the main excitation mechanism in the galaxies. The panels can be understood as, e.g. how galaxies with emission lines from AGN, LINERs or young stars are distributed as a function of $\Delta$PA. Almost all the late type galaxies in the sample are aligned, while early-type galaxies show a broader distribution in terms of $\Delta$PA. Most of the misaligned galaxies have AGN or LINER as their excitation mechanism. Excitation by star formation only, in early type galaxies tends to occur mostly in aligned $\Delta$PA $< 45^{\circ}$, with a smaller secondary peak in close to counter-rotating $\Delta$PA $\sim 180^{\circ}$ galaxies. The fraction of star forming early-type galaxies in the fourth bin ($135^{\circ}\leq\Delta$PA $\leq 180^{\circ}$) is marginally higher than the fraction of star forming early-type galaxies in the second and third bins at the 68\% and 95\% confidence level, respectively.}
	\label{FigA3}
\end{figure}

\begin{figure}[h]
	\centering
	\includegraphics[width=12cm]{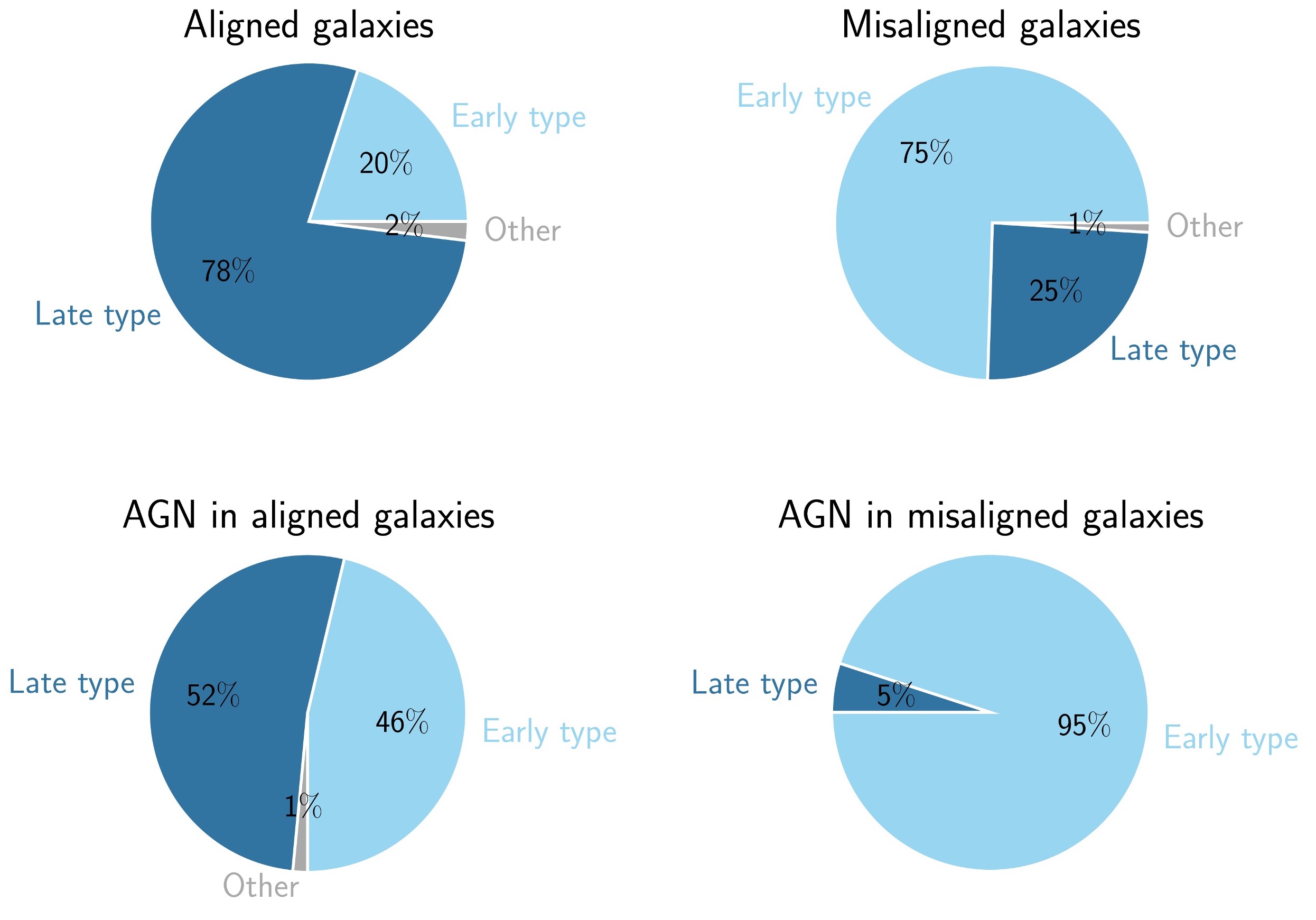}
	\caption{Pie charts showing the morphology classification of the sub-samples of galaxies. The top row shows the sub-sample of galaxies with an excitation classification, divided into `aligned' ($0^{\circ} \leq \Delta$PA $< 45^{\circ}$ - top left panel) and `misaligned' (45$^{\circ} \leq \Delta$PA $\leq 180^{\circ}$ - top right panel). The bottom row shows the sub-sample of galaxies classified as AGN, divided into AGN in `aligned' galaxies (bottom left panel) and AGN in `misaligned' galaxies (bottom right panel). The morphological classification is divided into early-type and late-type galaxies with `Other' referring to an unknown morphological classification. 75\% of all misaligned galaxies are in early-type galaxies while 95\% of AGN with misaligned hosts are in early-type galaxies.}
	\label{FigA4}
\end{figure}

\begin{figure*}
	\centering
	\hspace{-0.6cm}\includegraphics[width=12.5cm]{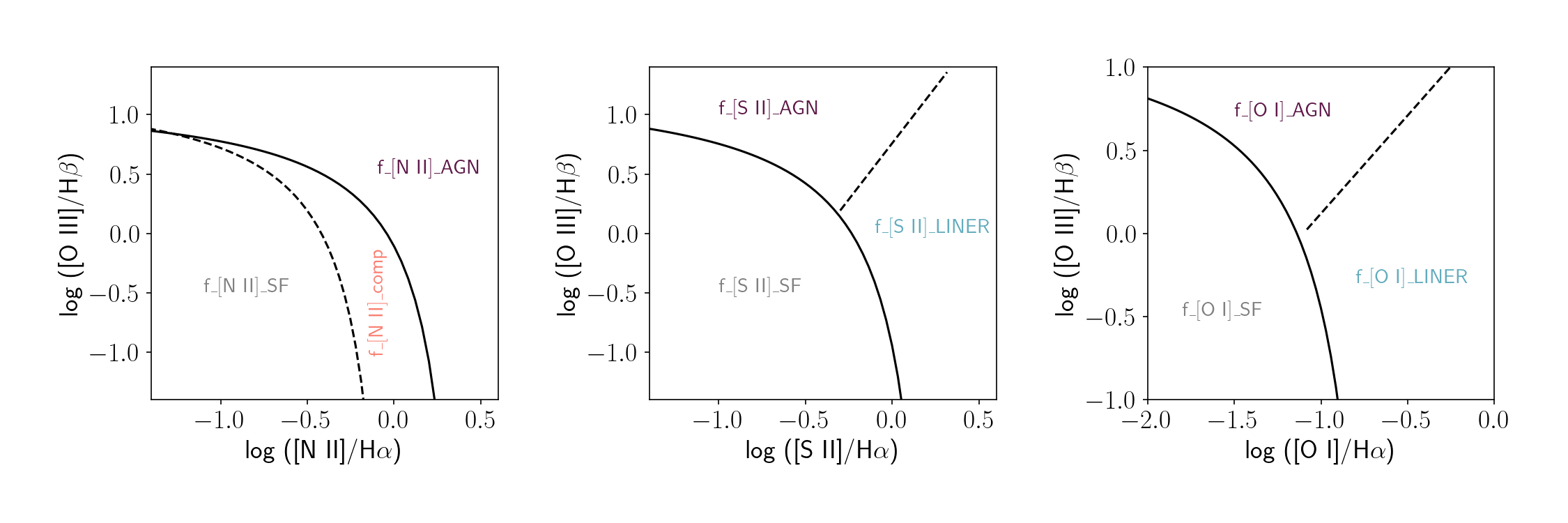}
	\caption{Illustration of the emission line diagrams used in this work and the theoretical regions of \cite{kewley06} indicated by the solid and dashed lines. The text labels refer to the fraction of spaxels that fall in each of the different regions of the diagrams. For clarity we use the same names in the figure as in the text (see \nameref{sec:methods}). The suffixes `SF' stand for star-forming regions, `AGN' for AGN-excitation regions, `comp' for composite regions (likely a combination of excitation by young stars and AGN) and 'LINER' for low-ionization nuclear emission-line region. The suffixes [N II], [S II] and [O I] refer to each of the corresponding BPT diagrams: $[$N II$]$/H$\alpha$, $[$S II$]$/H$\alpha$ and $[$O I$]$/H$\alpha$.}
	\label{FigA5}
\end{figure*}

\begin{figure*}
	\centering
	\includegraphics[width=12cm]{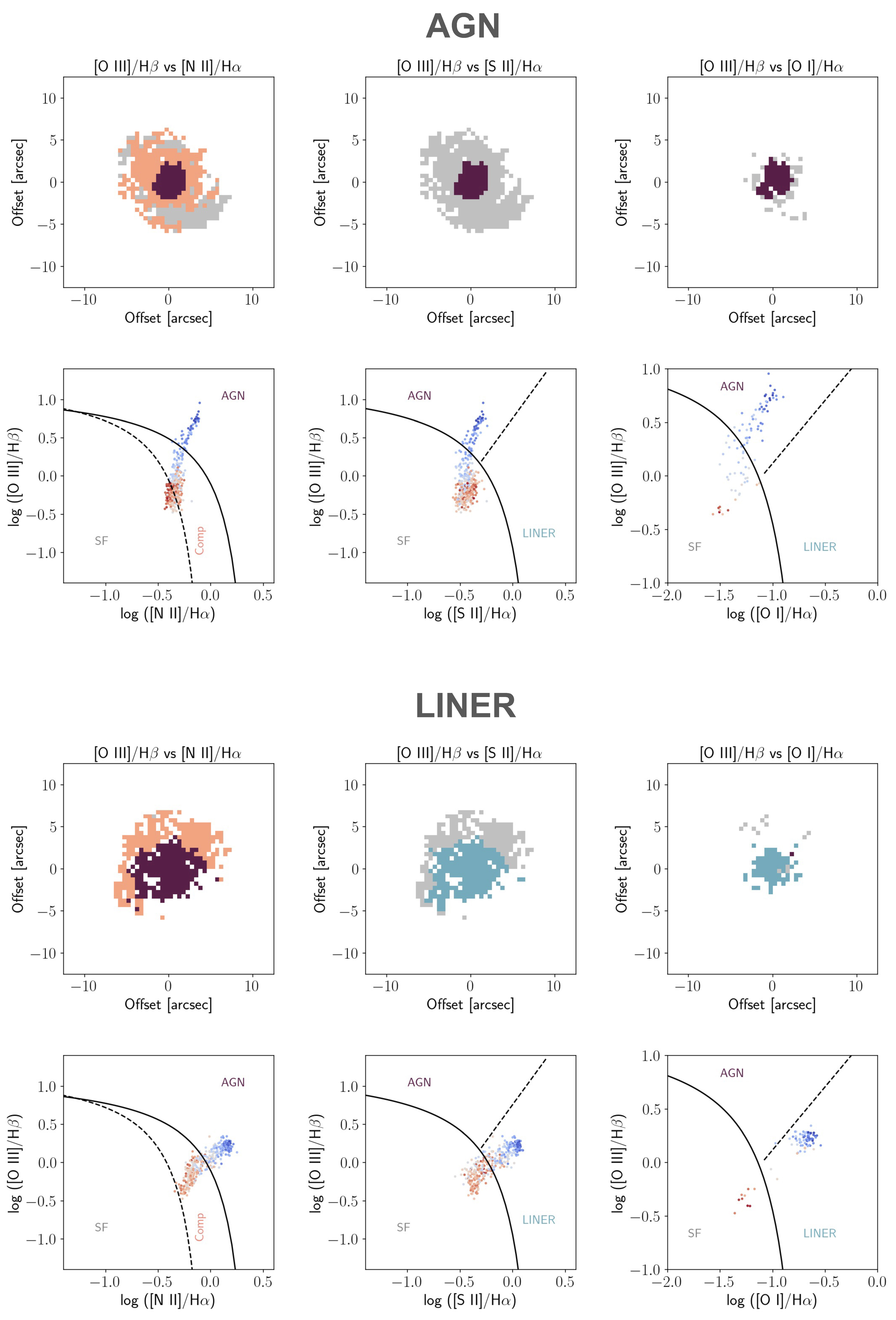}\\
	\caption{(Continues)}
\end{figure*}

\begin{figure*}\ContinuedFloat
	\caption{(Continued) Illustration of the line diagnostics used. Each figure shows the BPT analysis used to identify the excitation mechanisms: The three top panels show the galaxy spatial maps colour-coded by dominant excitation mechanism, purple for AGN, salmon for composite, grey for star formation and blue for LINER. Bottom panels show the BPT diagrams for each spaxel in the image, colour-coded as a function of distance from centre of the galaxy (shown as position (0,0) in the top panels). Blue symbols correspond to spaxels closest to the nucleus, red symbols to spaxels further away from the nucleus, with the range of colours calibrated for each individual galaxy. For example, the bluest point for each galaxy will be the spaxel with detected emission that is closest to its nucleus. That could be the central spaxel (distance = 0) or a spaxel that is further away from the nucleus if no emission is detected in the central spaxels. The solid and dashed lines are the classification boundaries from \cite{kewley06}. Each column refers for a specific line ratio, from left to right: $[$O III$]$/H$\beta$ vs $[$N II$]$/H$\alpha$, $[$O III$]$/H$\beta$ vs $[$S II$]$/H$\alpha$ and $[$O III$]$/H$\beta$ vs $[$O I$]$/H$\alpha$. The labels in the bottom row refer to the excitation classification in each region of the diagrams: AGN, LINER, SF (star-forming galaxies) or Comp (composite regions). The two figures (from top to bottom) show an example of a galaxy classified as AGN and another as LINER. An example of a star-forming galaxy is shown in Supplementary Information.}
	\label{FigA6}
\end{figure*}

\begin{figure}[h]
	\centering
	\includegraphics[width=12cm]{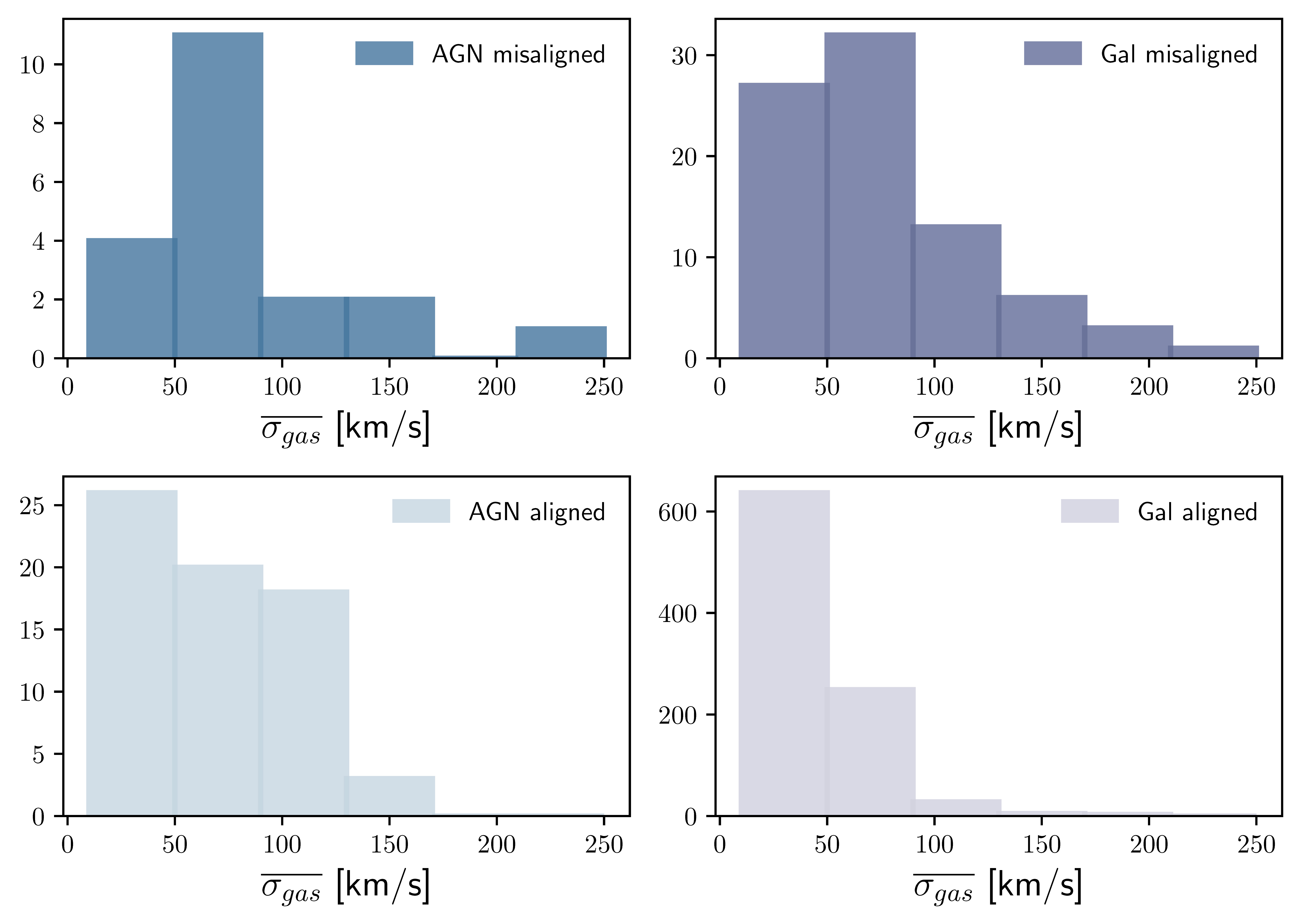}
	\caption{Histograms of the distribution of average gas velocity dispersion. The y-axis shows the number of galaxies per bin. The dark-coloured histograms at the top show misaligned galaxies while the light-coloured histograms at the bottom show the aligned galaxies. The distribution for AGN are shown in the left panels while the distribution for non-AGN galaxies are shown in the right panels.}
	\label{FigA7}
\end{figure}

\begin{figure}[h]
	\centering
	\includegraphics[width=12cm]{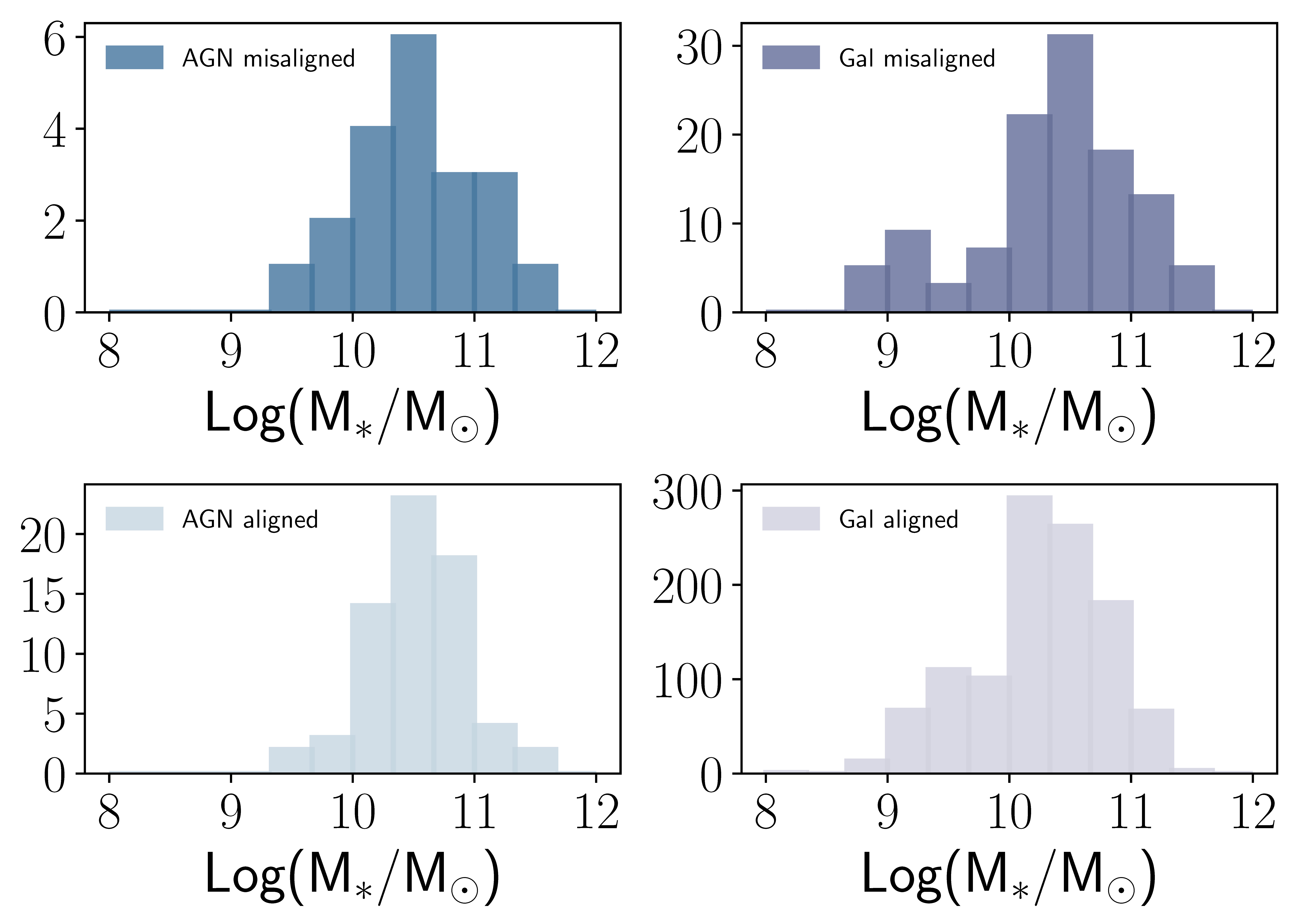}
	\caption{Histograms of the distribution of stellar mass in units of solar masses. The y-axis shows the number of galaxies per bin. The dark-coloured histograms at the top show misaligned galaxies while the light-coloured histograms at the bottom show the aligned galaxies. The distribution for AGN are shown in the left panels while the distribution for non-AGN galaxies are shown in the right panels.}
	\label{FigA8}
\end{figure}

\clearpage
\includepdf[pages=-]{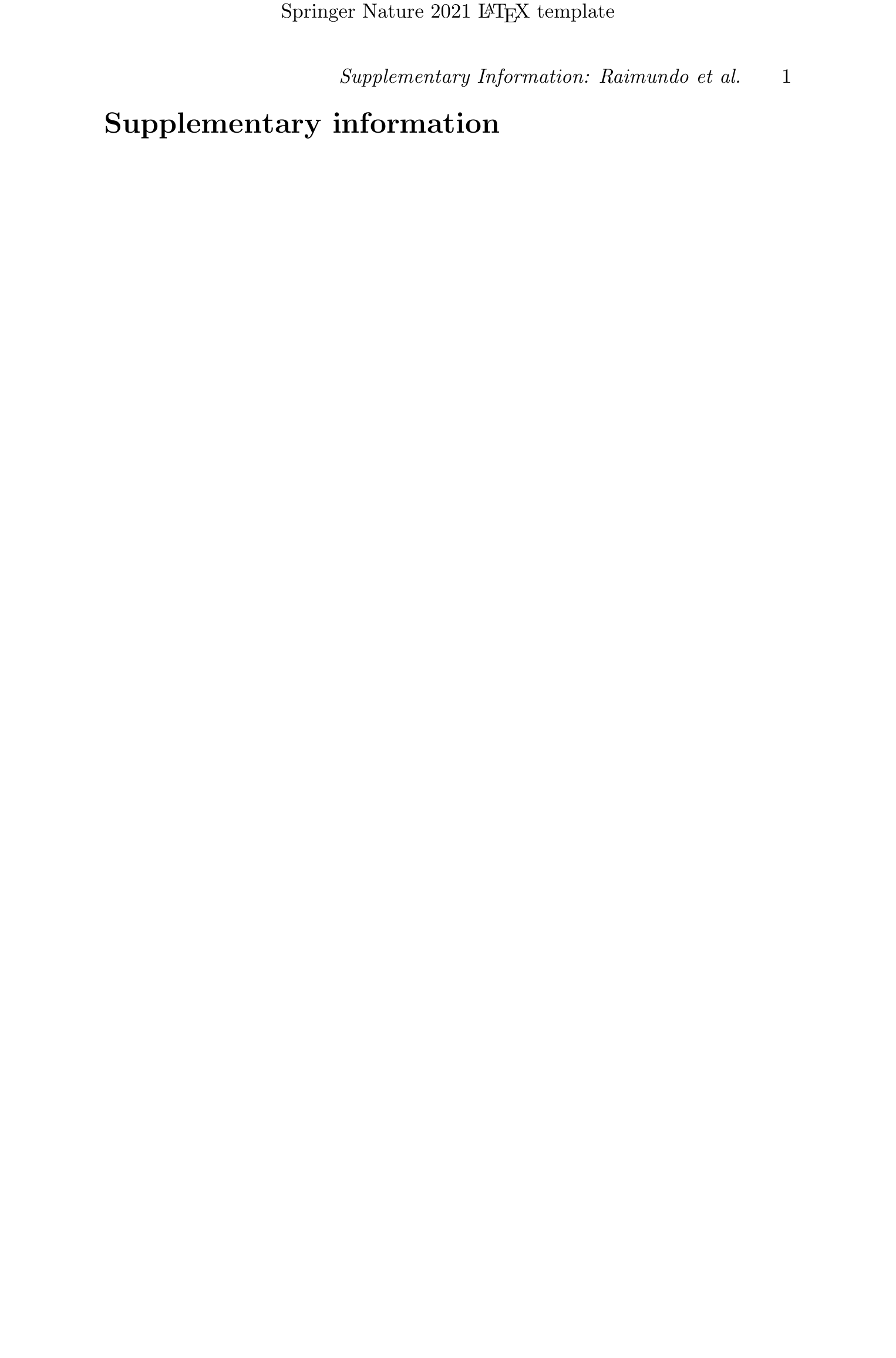}

\end{document}